\documentclass{revtex4}
\usepackage{makeidx}
\usepackage{eurosym}
\usepackage[utf8]{inputenc}
\usepackage{amsmath}
\usepackage{amsfonts}
\usepackage{t4phonet}
\usepackage{amssymb}
\usepackage{setspace}
\usepackage{tipa}
\usepackage{epsfig}
\usepackage{graphicx}
\usepackage{float}
\usepackage{subfigure}
\usepackage{xcolor}

\begin{document}

\title{Solitons in $\mathcal{PT}$-symmetric systems with spin-orbit coupling
and critical nonlinearity}
\author{Gennadiy Burlak$^1$, Zhaopin Chen$^2$\thanks{%
Corresponding author.}}
\email{zhaopin.chen@campus.technion.ac.il}
\author{Boris A. Malomed$^{3,4}$}
\affiliation{$^1$Centro de Investigaci\'{o}n en Ingenier\'{\i}a y Ciencias Aplicadas,
Universidad Aut\'{o}noma del Estado de Morelos, Cuernavaca, Morelos, Mexico\\
$^2$Physics Department and Solid State Institute, Technion, Haifa 32000,
Israel\\
$^{3}$Department of Physical Electronics, School of Electrical Engineering,
Faculty of Engineering, and Center for Light-Matter Interaction, Tel Aviv
University, P.O.B. 39040, Tel Aviv, Israel\\
$^4$Instituto de Alta Investigaci\'{o}n, Universidad de Tarapac\'{a},
Casilla 7D, Arica, Chile}

\begin{abstract}
We construct families of one-dimensional (1D) stable solitons in
two-component $\mathcal{PT}$-symmetric systems with spin-orbit coupling
(SOC) and quintic nonlinearity, which plays the critical role in 1D setups. The
system models light propagation in a dual-core waveguide with skewed
coupling between the cores. Stability regions for the solitons are
identified in the system's parameter space. They include the main
semi-infinite gap, and an additional finite \textit{annex gap}. Stability
boundaries are identified by means of simulations of the perturbed
evolution, which agree with results produced by the linear-stability
analysis for small perturbations. Distinct evolution scenarios are
identified for unstable solitons. Generally, they suffer blowup or decay,
while weakly unstable solitons transform into breathers. Due to a
regularizing effect of SOC, stationary solitons are also found beyond the
exceptional point, at which the $\mathcal{PT}$ symmetry breaks down, but
they are unstable. Interactions between adjacent solitons are explored too,
featuring rebound or merger followed by blowup. Slowly moving (tilted)
solitons develop weak oscillations, while fast ones are completely unstable.
Also considered is the reduced diffractionless system, which creates only
unstable solitons.
\end{abstract}

\maketitle

\section{Introduction}

The current work in the field of optics has drawn a great deal of interest
to using photonic media for emulation of various effects known in
condensed-matter and quantum physics, where direct experimental and
theoretical studies of such effects may be much more challenging. In many
cases, the photonic emulation is facilitated by the fact that the universal
equation of Schr\"{o}dinger type, which governs the paraxial light
propagation in linear and nonlinear media, is quite similar to the
fundamental Schr\"{o}dinger equation in quantum systems \cite%
{Longhi-similarity}. The similarity may be also established between
Hamiltonians of the condensed-matter or quantum settings and effective
Hamiltonians modeling optical phenomena.

Well-known examples of the emulation of diverse physical phenomenology by
photonics are provided by the Hall effect \cite{Hall}, topological
insulators \cite{insulator}, black holes \cite{black}, $\mathcal{PT}$
(parity-time) symmetry \cite{Bender}, which was realized theoretically \cite%
{Muga}-\cite{Zezyulin} and experimentally \cite{Morandotti,Segev,Peschel} in
diverse optical setups, and (pseudo-) spin-orbit coupling (SOC) \cite%
{Bliokh,KK,HS_BAM:2016}. Photonic SOC schemes were designed to simulate SOC
in atomic Bose-Einstein condensates (BECs) \cite{Spielman}-\cite{EPL},
which, in turn, was devised as the emulation of the SOC effect per se, which
plays a major role in physics of semiconductors \cite{spintronics,Rashba}.

$\mathcal{PT}$-symmetric schemes are built as ones which include
symmetrically placed and mutually balanced gain and loss elements, that
makes it possible to produce stable excitation spectra \cite{Bender},
provided that the strength of the gain-loss terms does not exceed a certain
critical value, which is often called the exceptional point \cite%
{exceptional1,exceptional,exceptional2}. In particular, $\mathcal{PT}$
symmetry can be realized in \textit{couplers}, i.e., dual-core optical
waveguides with linear coupling between the cores \cite%
{Wright:1989,Wabnitz2,Peng}. In this case, one core of the $\mathcal{PT}$%
-symmetric coupler carries the gain, while the mate one provides the
balancing loss \cite%
{Driben,Driben2,Alex:2012,BurMal:2013,Chen:2014,BurMal:2016,Fan}.

As concerns SOC in BEC, it is modeled by systems of Gross-Pitaevskii
equations for a spinor (two-component) wave function, coupled by linear
terms with first spatial derivatives \cite{Galitski,Goldman,Zhai}.
Accordingly, photonic emulation of such setups may be provided by an optical
coupler, in which amplitudes of the electromagnetic fields in the two
parallel cores correspond to the components of the BEC spinor wave function
\cite{KK,HS_BAM:2016,EZB:2020}. The aforementioned first derivatives are
then provided either by temporal dispersion of the inter-core coupling
constant in spatiotemporal optical couplers \cite{KK}, or by spatial shear
between two cores of the dual waveguide (its \textit{skewness}) \cite%
{HS_BAM:2016,EZB:2020}

Both $\mathcal{PT}$ symmetry and SOC being linear phenomena, it is natural
to consider their interplay. Dual-core waveguides, maintaining these
phenomena in essentially the same system, offer an optical platform for
integrating them. Furthermore, the same setup allows one to add intrinsic
nonlinearity of the waveguiding cores to the system, which opens the way to
construct solitons and consider other nonlinear effects \cite{HS_BAM:2016}.
In particular, it is especially interesting to consider the case of the
critical nonlinearity, which may give rise to the \textit{critical collapse}%
. It occurs in two- or one-dimensional (2D or 1D) nonlinear Schr\"{o}%
dinger/Gross-Pitaevskii equations with the cubic or quintic self-focusing,
respectively \cite{Berge,Sulem,Fibich}. Accordingly, solitons produced by
these equations, i.e., 2D \textit{Townes solitons} \cite{Townes} and their
1D counterparts \cite{AbdSal,Nakkeeran}, are unstable solutions (stability
of multidimensional solitons, including ones with embedded vorticity, may be
provided by the quadratic, i.e., second-harmonic-generating, nonlinearity
\cite{Mihalache}). In line with these well-known results, it was commonly
believed that 2D systems with cubic self-focusing in free space always
produce unstable solitons \cite{Sherman}. Nevertheless, in work \cite{Ben Li}
it was demonstrated that the linear SOC terms, added to the 2D system, make
it possible to produce completely stable solitons of two types, \textit{viz}%
., semi-vortices and mixed modes, which play the role of the system's ground
state. Then, it was demonstrated that the emulation of SOC in the 1D
dual-core coupler with the quintic self-focusing creates stable solitons,
instead of collapsing ones, in this case as well \cite{EZB:2020}.

The interplay of the $\mathcal{PT}$ symmetry and critical (cubic)
self-focusing in the presence of SOC in the 2D coupler, studied in the
presence of the optically-emulated SOC \cite{HS_BAM:2016}, is quite
interesting because, while the combination of the $\mathcal{PT}$-symmetric
gain-loss terms and critical nonlinearity makes the solitons fragile states,
the SOC terms secure their stability. The objective of the present work is
to address a similar problem in 1D, i.e., in the system of nonlinear-Schr%
\"{o}dinger equations modeling the dual-core optical waveguide with the
gain, loss, and quintic self-focusing in the two cores, which are coupled by
the above-mentioned skewed linear terms. The result is that vast families of
\textit{stable solitons} exist in this system. Boundaries of the stability
areas are identified in the system's parameter space.

The subsequent presentation is arranged as follows. The model of the
nonlinear $\mathcal{PT}$-symmetric coupler, which provides the optical
emulation of SOC, is formulated in Section II. It includes both the full
system and a reduced one, which neglects the terms representing the paraxial
diffraction in the dual-core coupler. The same section presents linearized
equations necessary for the analysis of the solitons' stability, and an
approximate analytical solution for the solitons in the case when both the $%
\mathcal{PT}$ symmetry and SOC may be treated as weak perturbations.
Numerical results are reported in Section III. They include stability areas
for solitons in the system's parameter space and results of systematic
numerical simulations for the evolution of unstable solitons (in particular,
weakly unstable solitons may avoid the collapse, transforming, instead, into
breathers or tilted (slowly moving) modes). Unlike the full system, the
reduced one produces solely unstable solitons. Interactions between stable
solitons are also studied by means of systematic simulations. The paper is
concluded by Section IV.

\section{The models}

Generalizing the considerations presented in Refs. \cite{KK,HS_BAM:2016} and
\cite{EZB:2020}, the interplay of the (pseudo-) SOC, critical nonlinearity,
and $\mathcal{PT}$ symmetry in the 1D setting may be realized in the planar
dual-core waveguide modeled by the following system of linearly-coupled
nonlinear Schr\"{o}dinger equations for complex amplitudes $u(x,z)$ and $%
v(z,x)$ of optical fields in the two cores:
\begin{eqnarray}
iu_{z}+\frac{1}{2}u_{xx}+|u|^{4}u+v-\delta \cdot v_{x} &=&i\gamma u,
\label{udelta} \\
iv_{z}+\frac{1}{2}v_{xx}+|v|^{4}v+u+\delta \cdot u_{x} &=&-i\gamma v.
\label{vdelta}
\end{eqnarray}%
Here $z$ and $x$ are, respectively, the scaled propagation distance and
transverse coordinate, coefficients of the paraxial diffraction, quintic
self-focusing, and straight inter-core coupling are scaled to be $1$, real $%
\delta >0$ represents the inter-core shear in the skewed coupler, and $%
\gamma >0$ is the strength of the $\mathcal{PT}$-symmetric gain and loss
terms.

In physical units, the $x=1$ and $z=1$ in Eqs. (\ref{udelta}) and (\ref%
{vdelta}) correspond to $\sim 50$ $\mathrm{\mu }$m and $1$ cm, respectively.
Assuming the use of optical materials which feature strong quintic
nonlinearity \cite{Cid}, an estimate of the power of laser beams which are
required to create the solitons considered below yields $\sim 10$ kW, the
respective power density being $\sim 10$ GW/cm$^{2}$.

In the absence of the gain and loss ($\gamma =0$), Eqs. (\ref{udelta}) and (%
\ref{vdelta}) conserve the total power (norm) and momentum of the wave field,%
\begin{equation}
P=\int_{-\infty }^{+\infty }\left( \left\vert u(x)\right\vert
^{2}+\left\vert v(x)\right\vert ^{2}\right) dx,~M=i\int_{-\infty }^{+\infty
}\left( u_{x}^{\ast }u+v_{x}^{\ast }v\right) dx.  \label{PM}
\end{equation}%
with $\ast $ standing for the complex conjugate. The $\mathcal{PT}$ terms
break the conservation, giving rise to the following evolution equations for
the power and momentum:%
\begin{eqnarray}
\frac{dP}{dz} &=&2\gamma \int_{-\infty }^{+\infty }\left(
|u|^{2}-|v|^{2}\right) dx,  \label{dP/dz} \\
\frac{dM}{dz} &=&2i\gamma \int_{-\infty }^{+\infty }\left( u_{x}^{\ast
}u-v_{x}^{\ast }v\right) dx.  \label{dM/dz}
\end{eqnarray}

Stationary solutions of Eqs. (\ref{udelta}) and (\ref{vdelta}) with real
propagation constant $k$ are looked as
\begin{equation}
\left\{ u,v\right\} =e^{ikz}\left\{ U(x),V(x)\right\} .  \label{uv}
\end{equation}%
The corresponding equations for complex functions $U(x)$ and $V(x)$ are%
\begin{eqnarray}
-kU+\frac{1}{2}\frac{d^{2}U}{dx^{2}}+|U|^{4}U+V-\delta \cdot \frac{dV}{dx}
&=&i\gamma U,  \label{Udelta} \\
-kV+\frac{1}{2}\frac{d^{2}V}{dx^{2}}+|V|^{4}V+U+\delta \cdot \frac{dU}{dx}
&=&-i\gamma V.  \label{Vdelta}
\end{eqnarray}%
\newline

It is relevant to identify the spectrum of the linearized system. Looking
for small-amplitude solutions to Eqs. (\ref{Udelta}) and (\ref{Vdelta}) in
the form of plane waves, \
\begin{equation}
\left\{ U,V\right\} \sim \exp \left( iqx\right) ,  \label{tail}
\end{equation}%
with wavenumber $q$, one derives the following dispersion relation between $%
k $ and $q^{2}$:%
\begin{equation}
k=-(1/2)q^{2}\pm \sqrt{1-\gamma ^{2}+\delta ^{2}q^{2}}.  \label{k}
\end{equation}%
The spectrum is real, i.e., the $\mathcal{PT}$ symmetry holds, under the
condition of $\gamma <1$, while $\gamma =1$ corresponds to the \textit{%
exceptional point} \cite{exceptional1,exceptional,exceptional2} of the $%
\mathcal{PT}$-symmetric system. In terms of the dual-core system, the
strengths of the straight inter-core coupling and gain-loss terms are
exactly equal at this point.

At $\gamma >1$, the $\mathcal{PT}$ symmetry breaks down, and spectrum (\ref%
{k}) becomes complex (unstable). It is relevant to note that, in the absence
of SOC (at $\delta =0$), the entire spectrum blows up (immediately becomes
complex) at any $\gamma >1$. On the other hand, SOC provides a regularizing
effect, as, at $\gamma >1$ and $\delta \neq 0$, spectrum (\ref{k}) is
complex only at sufficiently small wavenumbers, \textit{viz}., at
\begin{equation}
q^{2}<\left( \gamma ^{2}-1\right) /\delta ^{2}.  \label{complex}
\end{equation}%
It is shown below that, as a result of the regularization, Eqs. (\ref{Udelta}%
) and (\ref{Vdelta}) may produce stationary soliton solutions at $\gamma >1$%
, although those solutions are unstable, see Fig. \ref{fig10} below.

Inversion of Eq. (\ref{k}) yields $q^{2}$ as a function if $k$:%
\begin{equation}
q^{2}=2\left( \delta ^{2}-k\pm \sqrt{\delta ^{4}-2\delta ^{2}k+1-\gamma ^{2}}%
\right) .  \label{q}
\end{equation}%
It follows from Eq. (\ref{q}) that solitons may populate the \textit{%
semi-infinite bandgap} (SIG) of the system's spectrum, which is%
\begin{equation}
k>k_{\mathrm{SIG}}\equiv \frac{1}{2}\left( \delta ^{2}+\frac{1-\gamma ^{2}}{%
\delta ^{2}}\right) ,  \label{kmax1}
\end{equation}%
In the SIG, Eq. (\ref{q}) gives rise to complex values of $q$, which implies
that solitons, if they exist, feature exponentially decaying tails with
spatial oscillations, such as $\exp \left( -a|x|\right) \cdot \cos (bx)$,
with real constants $a>0$ and $b$. In addition to that, in the case of%
\begin{equation}
\delta ^{4}<1-\gamma ^{2},  \label{delta <}
\end{equation}%
there exists an extra finite \textit{annex gap} (AG):
\begin{equation}
\sqrt{1-\gamma ^{2}}<k<\frac{1}{2}\left( \delta ^{2}+\frac{1-\gamma ^{2}}{%
\delta ^{2}}\right) ,  \label{kmax2}
\end{equation}%
cf. its counterpart for $\gamma =0$ obtained in Ref. \cite{EZB:2020}. In
interval (\ref{kmax2}), Eq. (\ref{q}) yields purely imaginary $q$, hence the
solitons, if they populate the AG, have exponentially decaying tails without
oscillations.

On the other hand, under condition
\begin{equation}
\delta ^{4}>1-\gamma ^{2},  \label{opposite}
\end{equation}
SIG is the single bandgap, and the AG does not exist in this case.

\begin{figure}[tbp]
\centering{\includegraphics[width=3.5in]{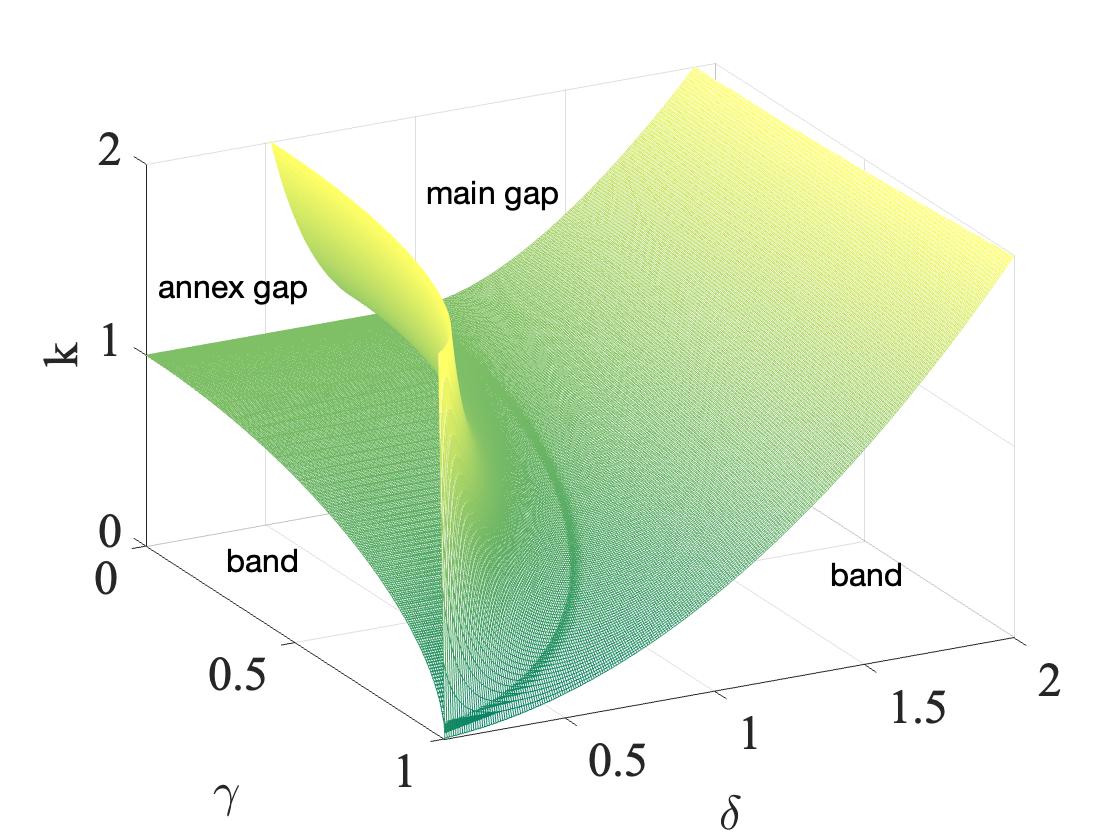}}
\caption{(Color online.) The bandgap structure plotted according to the
dispersion relation (\protect\ref{q}) and Eqs. (\protect\ref{kmax1})-(%
\protect\ref{kmax2}).}
\label{fig0}
\end{figure}

The region of the existence of the gaps can be also defined in the plane of $%
\left( \gamma ,\delta \right) $ for fixed $k$. Indeed, it follows from Eqs. (%
\ref{kmax1}) that the SIG exists in the area defined by the system of
inequalities%
\begin{equation}
1+\delta ^{4}-2k\delta ^{2}\equiv \gamma _{\mathrm{SIG}}^{2}<\gamma ^{2}<1,
\label{fixed_k_main}
\end{equation}%
In addition to it, the AG exists in the area defined by the following
inequalities:%
\begin{equation}
1-k^{2}<\gamma ^{2}<1-\delta ^{4}.  \label{fixed_k_AG}
\end{equation}%
%
%
%
At the exceptional point $\gamma =1$, the AG does not exist, while the main
SIG takes a simple form,%
\begin{equation}
k>\delta ^{2}/2.  \label{simple}
\end{equation}%
In Fig. \ref{fig0}, the bandgap structure as a whole is displayed in the
space of $\left( \delta ,\gamma ,k\right) $, see also Fig. \ref{fig1} below.

Furthermore, for \textquotedblleft moving" solitons (in fact, ones tilted in
the spatial domain) Eqs. (\ref{udelta}), (\ref{vdelta}) and (\ref{Udelta}), (%
\ref{Vdelta}) can be rewritten in terms of $z$ and the tilted coordinate,
\begin{equation}
\xi \equiv x-cz,  \label{xi}
\end{equation}%
where $c$ is the tilt (\textquotedblleft velocity"). The resulting
dispersion relation, including $c$, is rather cumbersome, but it takes a
simple form at the exceptional point ($\gamma =1$), at which the SIG becomes%
\begin{equation}
k>k_{\mathrm{SIG}}(\gamma =1,c)=\left( |c|+\delta \right) ^{2}/2,
\label{simple-c}
\end{equation}%
cf. Eq. (\ref{simple}). According to Eq. (\ref{simple-c}), the SIG decreases
with the increase of $|c|$.

\subsection{Approximate soliton solutions}

In the absence of the SOC terms ($\delta =0$), exact symmetric solitons
produced by Eqs. (\ref{Udelta}) and (\ref{Vdelta}) are obvious:%
\begin{eqnarray}
U_{\delta =0}(x) &=&\sqrt{\sqrt{1-\gamma ^{2}}-i\gamma }U_{0}(x;\gamma ),
\label{U0} \\
V_{\delta =0}(x) &=&\sqrt{\sqrt{1-\gamma ^{2}}+i\gamma }U_{0}(x;\gamma ),
\label{V0} \\
U_{0}(x;\gamma ) &=&\frac{\left[ 3\left( k-\sqrt{1-\gamma ^{2}}\right) %
\right] ^{1/4}}{\sqrt{\cosh \left( 2\sqrt{2\left( k-\sqrt{1-\gamma ^{2}}%
\right) }x\right) }}.  \label{00}
\end{eqnarray}%
They exist for $k>\sqrt{1-\gamma ^{2}}$, and are definitely unstable, for
the same reason as the usual 1D Townes solitons, corresponding to $\gamma =0$
\cite{AbdSal}.

When both $\gamma $ and $\delta $ are small parameters, an approximate
solution can be written as a straightforward extension of Eq. (43) from Ref.
\cite{EZB:2020}, where it was obtained, in a real form, for $\gamma =0$:%
\begin{eqnarray}
U(x) &\approx &\left( 1-i\frac{\gamma }{2}\right) U_{0}(x;0)-\frac{\delta }{2%
}\frac{dU_{0}(x;0)}{dx},  \label{pertU} \\
V(x) &\approx &\left( 1+i\frac{\gamma }{2}\right) U_{0}(x;0)+\frac{\delta }{2%
}\frac{dU_{0}(x;0)}{dx}.  \label{pertV}
\end{eqnarray}%
Here, $U_{0}$ is the expression given by Eq. (\ref{00}).

\subsection{Equations for small perturbations}

For the study of stability of solitons, perturbed solutions with complex
eigenmodes of small perturbations, $\phi _{1,2}(x)$ and $\psi _{1,2}(x)$,
are introduced as%
\begin{eqnarray}
u &=&e^{ikz}\left[ U(x)+\exp \left( \sigma z\right) \phi _{1}(x)+\exp \left(
\sigma ^{\ast }z\right) \phi _{2}^{\ast }(x)\right] ,  \label{upert} \\
v &=&e^{ikz}\left[ V(x)+\exp \left( \sigma z\right) \psi _{1}(x)+\exp \left(
\sigma ^{\ast }z\right) \psi _{2}^{\ast }(x)\right] ,  \label{vpert}
\end{eqnarray}%
where $U(x)$ and $V(x)$ represent the unperturbed solution, and $\sigma $ is
the instability growth rate (it may be complex). The resulting problem for
eigenmodes amounts to the linearized system of equations derived by the
substitution of expressions (\ref{upert}) and (\ref{vpert}) in Eqs. (\ref%
{udelta}) and (\ref{vdelta}):%
\begin{eqnarray}
&&\left( -k+i\sigma -i\gamma \right) \phi _{1}+\frac{1}{2}\frac{d^{2}\phi
_{1}}{dx^{2}}+3\left\vert U(x)\right\vert ^{4}\phi _{1}+2\left\vert
U(x)\right\vert ^{2}\left( U(x)\right) ^{2}\phi _{2}+\psi _{1}-\delta \frac{%
d\psi _{1}}{dx}=0,  \notag \\
&&\left( -k-i\sigma +i\gamma \right) \phi _{2}+\frac{1}{2}\frac{d^{2}\phi
_{2}}{dx^{2}}+3\left\vert U(x)\right\vert ^{4}\phi _{2}+2\left\vert
U(x)\right\vert ^{2}\left( U^{\ast }(x)\right) ^{2}\phi _{1}+\psi
_{2}-\delta \frac{d\psi _{2}}{dx}=0,  \notag \\
&&  \label{BdG} \\
&&\left( -k+i\sigma +i\gamma \right) \psi _{1}+\frac{1}{2}\frac{d^{2}\psi
_{1}}{dx^{2}}+3\left\vert V(x)\right\vert ^{4}\psi _{1}+2\left\vert
V(x)\right\vert ^{2}\left( V(x)\right) ^{2}\psi _{2}+\phi _{1}+\delta \frac{%
d\phi _{1}}{dx}=0,  \notag \\
&&\left( -k-i\sigma -i\gamma \right) \psi _{2}+\frac{1}{2}\frac{d^{2}\psi
_{2}}{dx^{2}}+3\left\vert V(x)\right\vert ^{4}\psi _{2}+2\left\vert
V(x)\right\vert ^{2}\left( V^{\ast }(x)\right) ^{2}\psi _{2}+\phi
_{2}+\delta \frac{d\phi _{2}}{dx}=0.  \notag
\end{eqnarray}%
As usual, the solitons are stable if all eigenvalues $\sigma $ have Re$%
\left( \sigma \right) \leq 0$.

\subsection{The reduced system}

Following Refs. \cite{HS_BAM:2016} and \cite{EZB:2020}, it is interesting to
consider the reduced version of the system for broad solitons, in which the
diffraction terms (second derivatives) may be omitted. In this case,
rescaling makes it possible to fix $\delta \equiv 1$, and Eqs. (\ref{udelta}%
), (\ref{vdelta}) and (\ref{Udelta}), (\ref{Vdelta}) are replaced,
respectively, by%
\begin{eqnarray}
iu_{z}+|u|^{4}u+v-v_{x} &=&i\gamma u,  \label{u2} \\
iv_{z}+|v|^{4}v+u+u_{x} &=&-i\gamma v,  \label{v2}
\end{eqnarray}%
\begin{eqnarray}
-kU+|U|^{4}U+V-\frac{dV}{dx} &=&i\gamma U,  \label{U2} \\
-kV+|V|^{4}V+U+\frac{dU}{dx} &=&-i\gamma V.  \label{V2}
\end{eqnarray}%
In this case, dispersion relation (\ref{k}) is replaced by%
\begin{equation}
k=\pm \sqrt{1-\gamma ^{2}+q^{2}}.  \label{k2}
\end{equation}%
The condition of the $\mathcal{PT}$ symmetry keeps the same form as above, $%
\gamma <1$. In this case, Eq. (\ref{k2}) gives rise to a finite bandgap,
unlike the SIC generated by the full system,%
\begin{equation}
|k|<\sqrt{1-\gamma ^{2}}.  \label{gap}
\end{equation}%
Note that, at the exceptional point $\gamma =1$, dispersion relation (\ref%
{k2}) takes the form of the \textit{Dirac's cone}, $k=\pm |q|$, with bandgap
(\ref{gap}) shrinking to nil.

It is relevant to mention that, by means of substitution $u=A+iB,v=iA+B$,
the linear version of Eqs. (\ref{u2}) and (\ref{v2}) can be transformed into
the spinor $\mathcal{PT}$-symmetric system for fileds $A$ and $B$, which was
introduced in Ref. \cite{Barash1}.

In the absence of the gain and loss ($\gamma =0$), it is straightforward to
find exact real solutions to Eqs. (\ref{U2}) and (\ref{V2}), extending the
method elaborated in Ref. \cite{HS_BAM:2016} for effectively the same
system, but with the cubic nonlinearity. The exact solution is%
\begin{gather}
\left\{ U(x),V(x)\right\} =A(x)\left\{ \cos \theta (x),\sin \theta
(x)\right\} ,  \notag \\
A^{4}(x)=12\frac{k-\sin (2\theta (x))}{4-3\sin ^{2}(2\theta (x))},
\label{exact} \\
\theta (x)=-\frac{\pi }{4}+\arctan \left[ \frac{1+k}{\sqrt{1-k^{2}}}\tanh
\left( 2\sqrt{1-k^{2}}x\right) \right] .  \notag
\end{gather}%
In Ref. \cite{EZB:2020} it was produced only in an approximate form.
Although the solution (\ref{exact}) is completely unstable in the framework
of Eqs. (\ref{u2}) and (\ref{v2}), the instability is weak, i.e., the
solution is a physically relevant one, for $0<1+k\ll 1$ \cite{EZB:2020}.

As concerns the \textquotedblleft moving" (tilted) solutions, for which Eqs.
(\ref{u2}), (\ref{v2}) and (\ref{U2}), (\ref{V2}) should be rewritten in
terms of $z$ and moving coordinate (\ref{xi}), the accordingly modified
dispersion relation (\ref{k2}) is%
\begin{equation}
k=-cq\pm \sqrt{1-\gamma ^{2}+q^{2}}.  \label{k3}
\end{equation}%
It gives rise to a narrower gap, in comparison with one (\ref{gap}):%
\begin{equation}
|k|<\sqrt{\left( 1-c^{2}\right) \left( 1-\gamma ^{2}\right) },  \label{gap-c}
\end{equation}%
provided that $c^{2}<1$. In the case of $c^{2}>1$, gap (\ref{gap-c}) does
not exist.

The stability of stationary solutions to Eqs. (\ref{u2}) and (\ref{v2}) can
be explored using the same ansatz (\ref{upert}), (\ref{vpert}) as introduced
above. The corresponding linearized equations for the perturbation
eigenmodes are obtained from Eqs. (\ref{BdG}) by dropping the second
derivatives:
\begin{eqnarray}
\left( -k+i\sigma -i\gamma \right) \phi _{1}+3\left\vert U(x)\right\vert
^{4}\phi _{1}+2\left\vert U(x)\right\vert ^{2}\left( U(x)\right) ^{2}\phi
_{2}+\psi _{1}-\frac{d\psi _{1}}{dx} &=&0,  \notag \\
\left( -k-i\sigma +i\gamma \right) \phi _{2}+3\left\vert U(x)\right\vert
^{4}\phi _{2}+2\left\vert U(x)\right\vert ^{2}\left( U^{\ast }(x)\right)
^{2}\phi _{1}+\psi _{2}-\frac{d\psi _{2}}{dx} &=&0,  \notag \\
&&  \label{BdG2} \\
\left( -k+i\sigma +i\gamma \right) \psi _{1}+3\left\vert V(x)\right\vert
^{4}\psi _{1}+2\left\vert V(x)\right\vert ^{2}\left( V(x)\right) ^{2}\psi
_{2}+\phi _{1}+\frac{d\phi _{1}}{dx} &=&0,  \notag \\
\left( -k-i\sigma -i\gamma \right) \psi _{2}+3\left\vert V(x)\right\vert
^{4}\psi _{2}+2\left\vert V(x)\right\vert ^{2}\left( V^{\ast }(x)\right)
^{2}\psi _{2}+\phi _{2}+\frac{d\phi _{2}}{dx} &=&0.  \notag
\end{eqnarray}

\section{Numerical results}

\subsection{The stability chart for soliton families}

Stationary soliton solutions of Eq. (\ref{Udelta}) and (\ref{Vdelta}) were
obtained by means of the squared-operator iteration method \cite%
{Yang:2008,Yang:2010}. Then, their stability was identified through the set
of eigenvalues produced by a numerical solution of linearized equations (\ref%
{BdG}), and verified by simulations of Eqs. (\ref{udelta}) and (\ref{vdelta}%
) for perturbed evolution of the solitons, using the split-step Fourier
method. The numerical solutions were constructed, chiefly, in the domain of
size $|x|\leq 20$, covered by a numerical mesh of $512$ sites, with
absorbing boundary conditions.

\begin{figure}[tbp]
\centering{\subfigure[]{\includegraphics[width=3in]{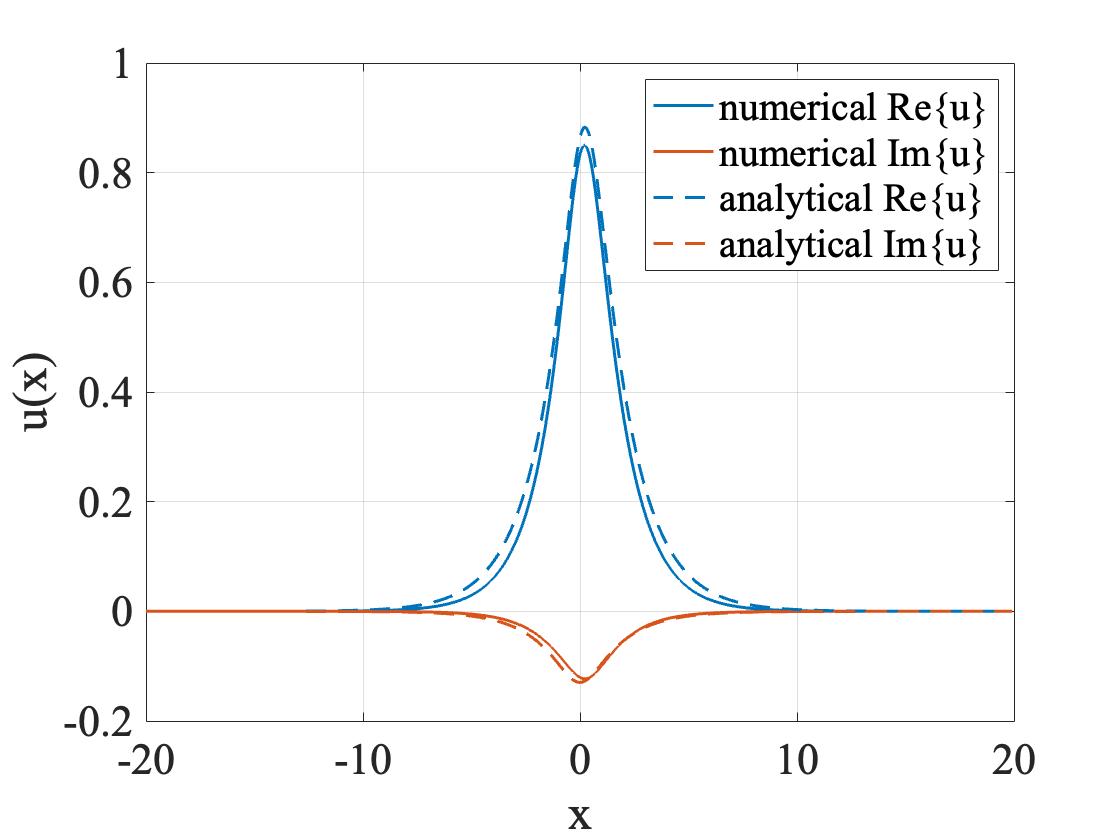}}
\subfigure[]{
        \includegraphics[width=3in]{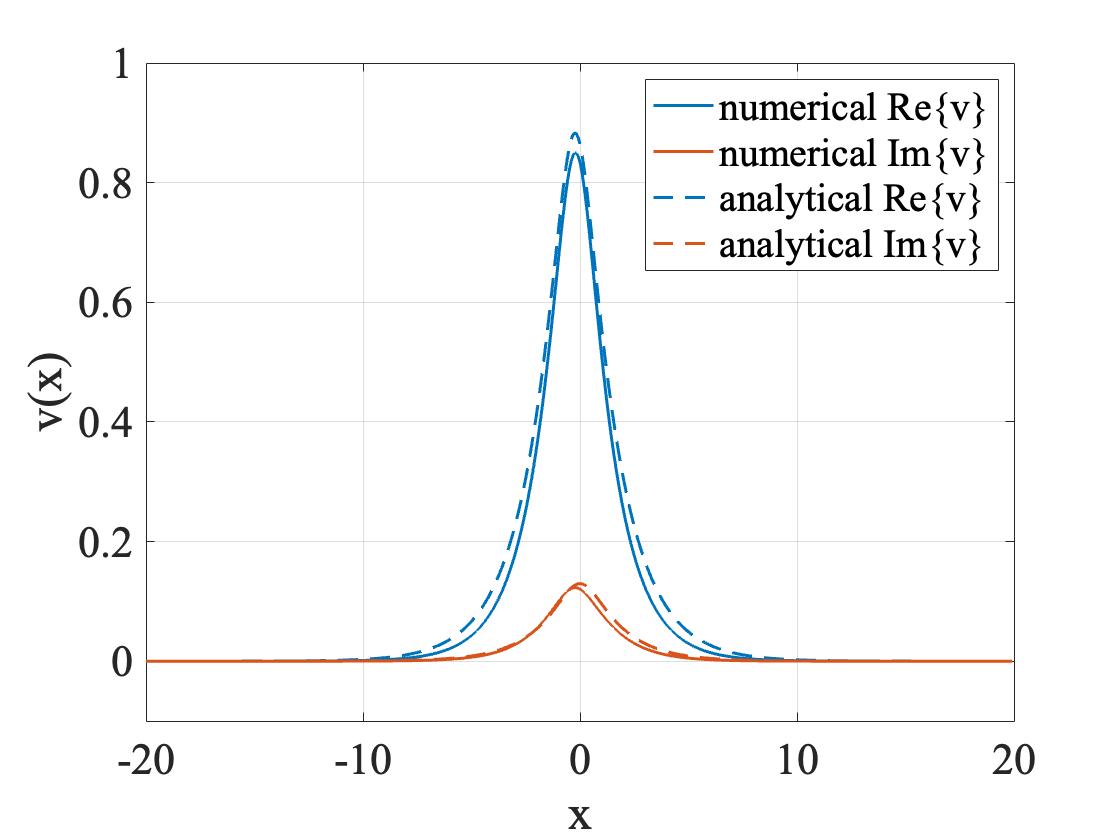}}} 
\caption{(Color online). Numerically found components of a stable soliton
with $\protect\delta =0.35,\protect\gamma =0.3$, and $k=1.14$, and their
counterparts produced by the analytical approximation based on Eqs. (\protect
\ref{pertU}) and (\protect\ref{pertV}).}
\label{fig6}
\end{figure}

\begin{figure}[tbp]
\centering{\includegraphics[width=3in]{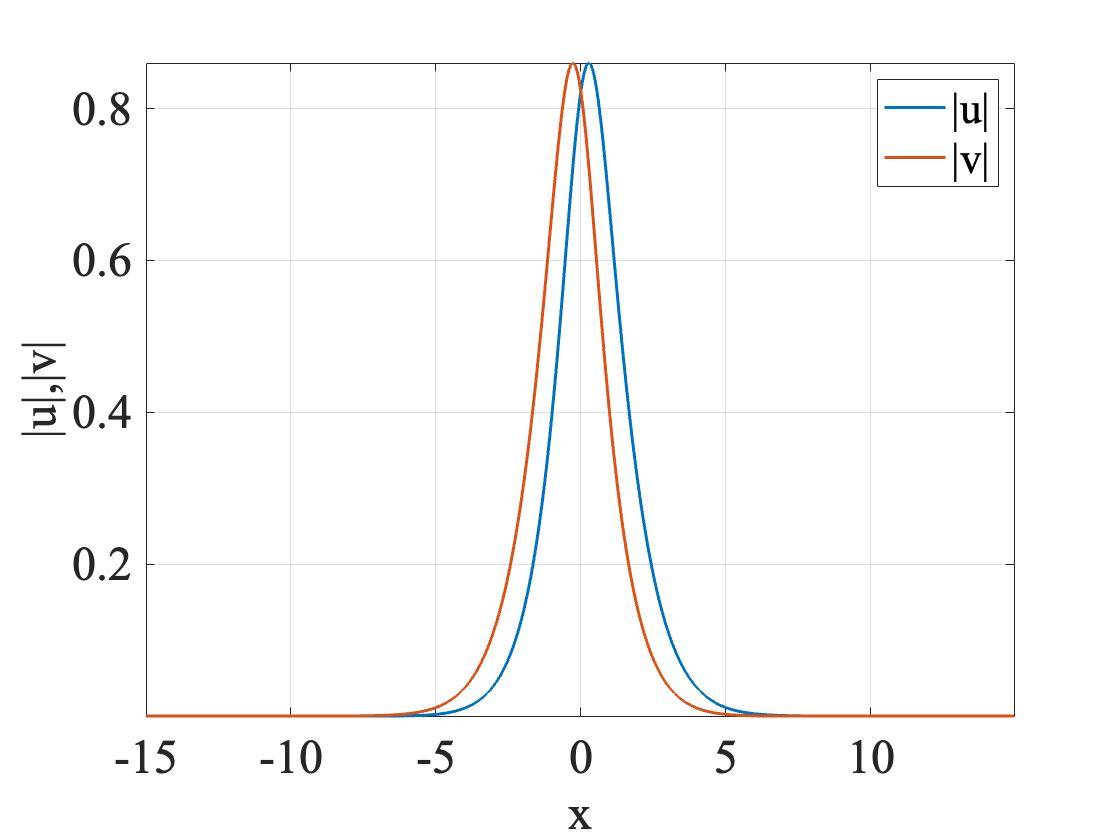}}
\caption{(Color online). A typical example of a stable soliton in the system
with $\protect\delta =0.8,\protect\gamma =0.1$, corresponding to the
propagation constant $k=1.2$. This soliton is stable in terms of eigenvalues
of small perturbations and direct simulations of the perturbed evolution.}
\label{fig2}
\end{figure}

\begin{figure}[tbp]
\centering{\ \subfigure[]{\includegraphics[width=3in]{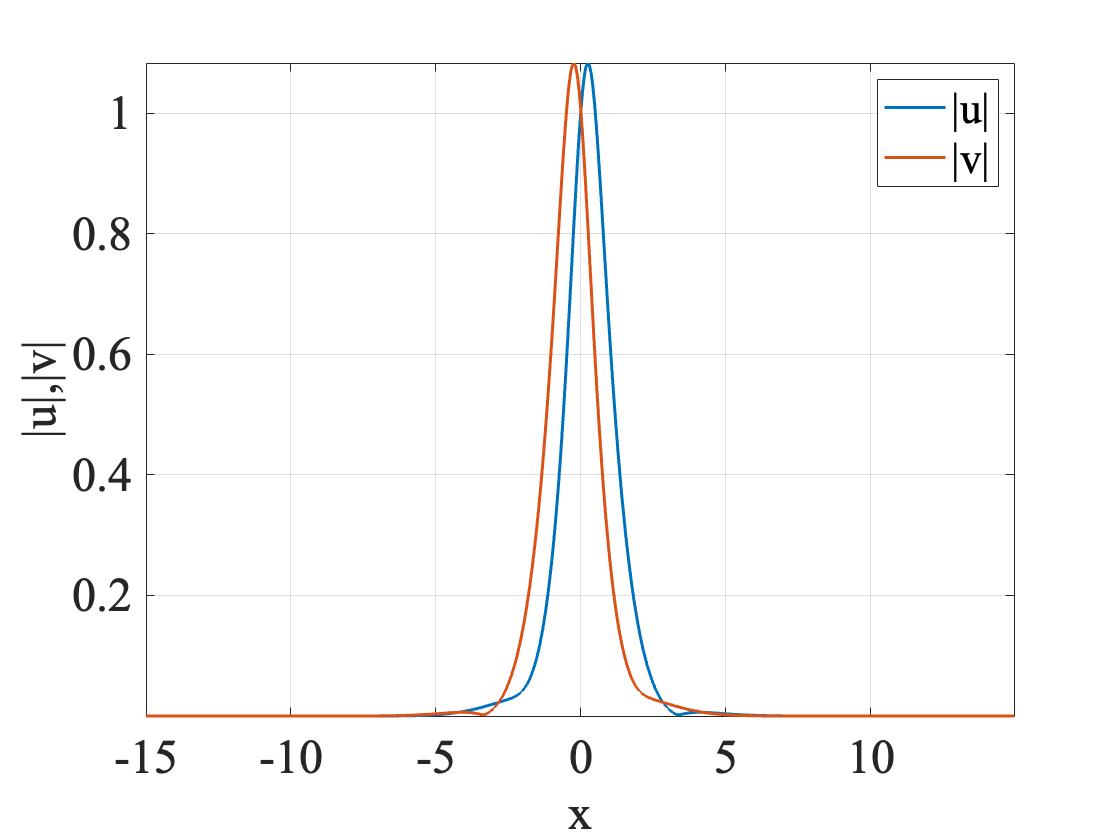}}
\subfigure[]{
        \includegraphics[width=3in]{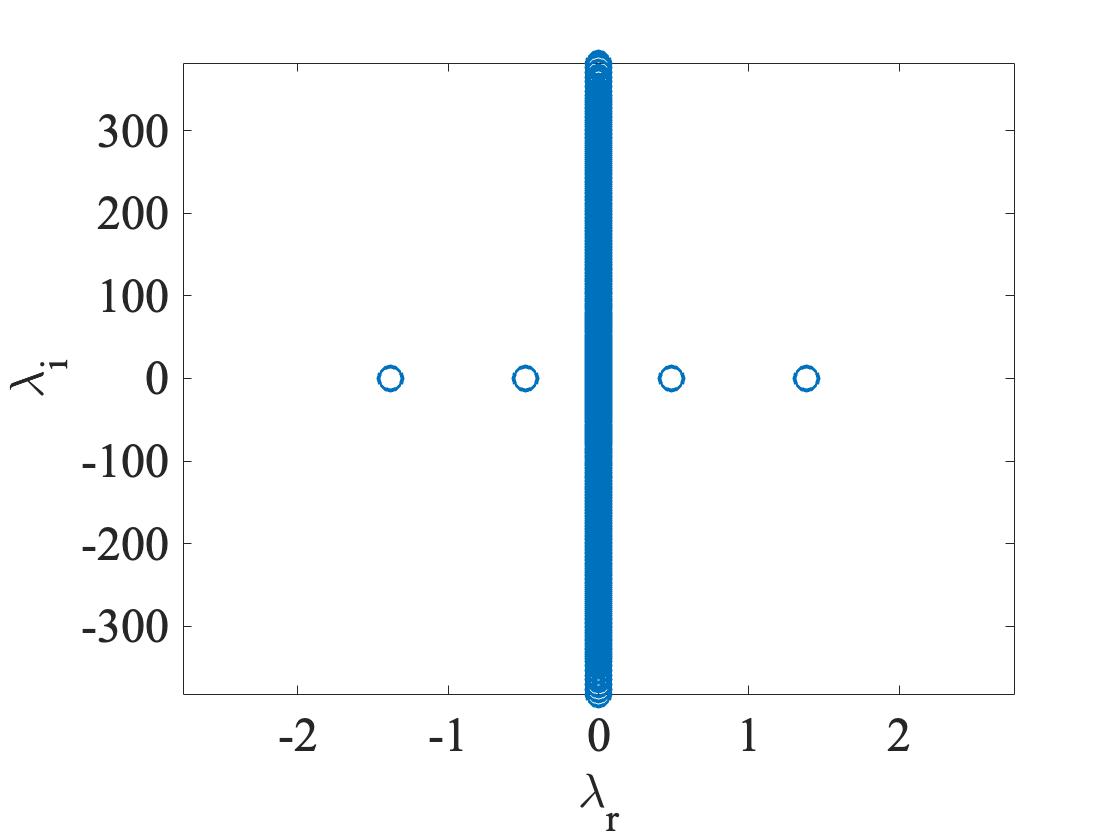}}
\subfigure[]{
        \includegraphics[width=3in]{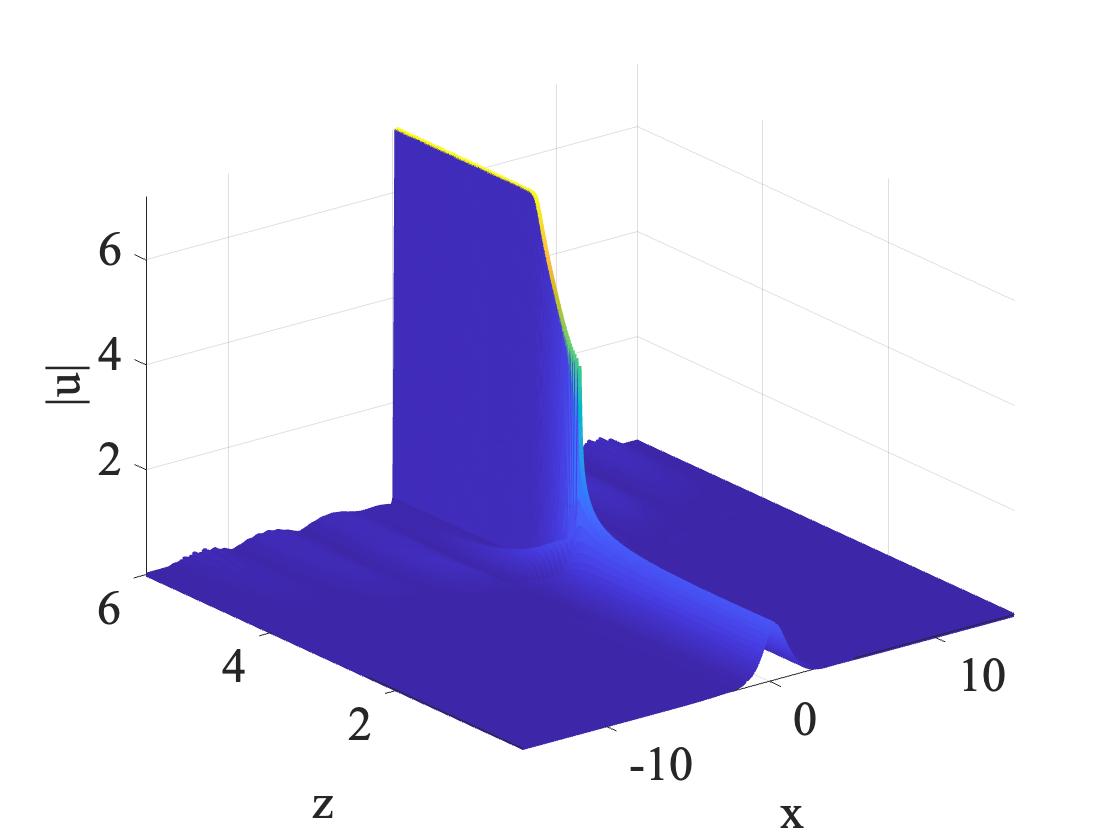}}
\subfigure[]{
        \includegraphics[width=3in]{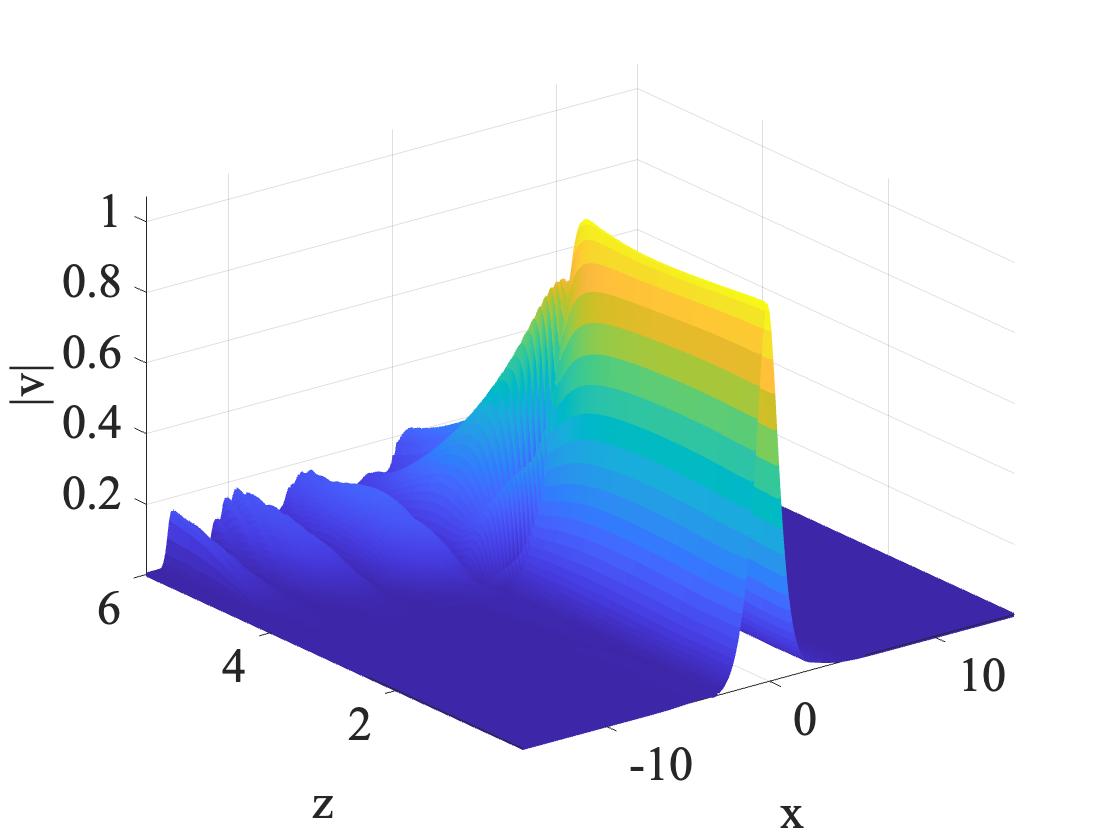}}}
\caption{(Color online). A typical example of an unstable soliton in the
system with $\protect\delta =0.8,\protect\gamma =0.8$ for $k=1.2$. (a) The
shape of the soliton; (b) the spectrum of eigenvalues for small
perturbations, including unstable ones; (c) and (d) the evolution of the
soliton under the action of a small random perturbation at the amplitude
level of $1\%$. Eventually, one component of the soliton blows up and the
other one decays. In (c), the growth of the amplitude is limited by the
numerical scheme.}
\label{fig3}
\end{figure}

First, Fig. \ref{fig6} shows good agreement of the analytical approximation,
given by Eqs. (\ref{pertU}) and (\ref{pertV}), with a numerical solution
obtained for a moderately small value of $\delta $ and small $\gamma $.
Additional examples of stable and unstable numerically found solitons are
shown in Figs. \ref{fig2} and \ref{fig3}. In particular, the analytical
approximation correctly predicts splitting $\Delta x\simeq \delta $ between
peaks of the linearly coupled components.

\begin{figure}[tbp]
\centering{\ \subfigure[]{\includegraphics[width=3in]{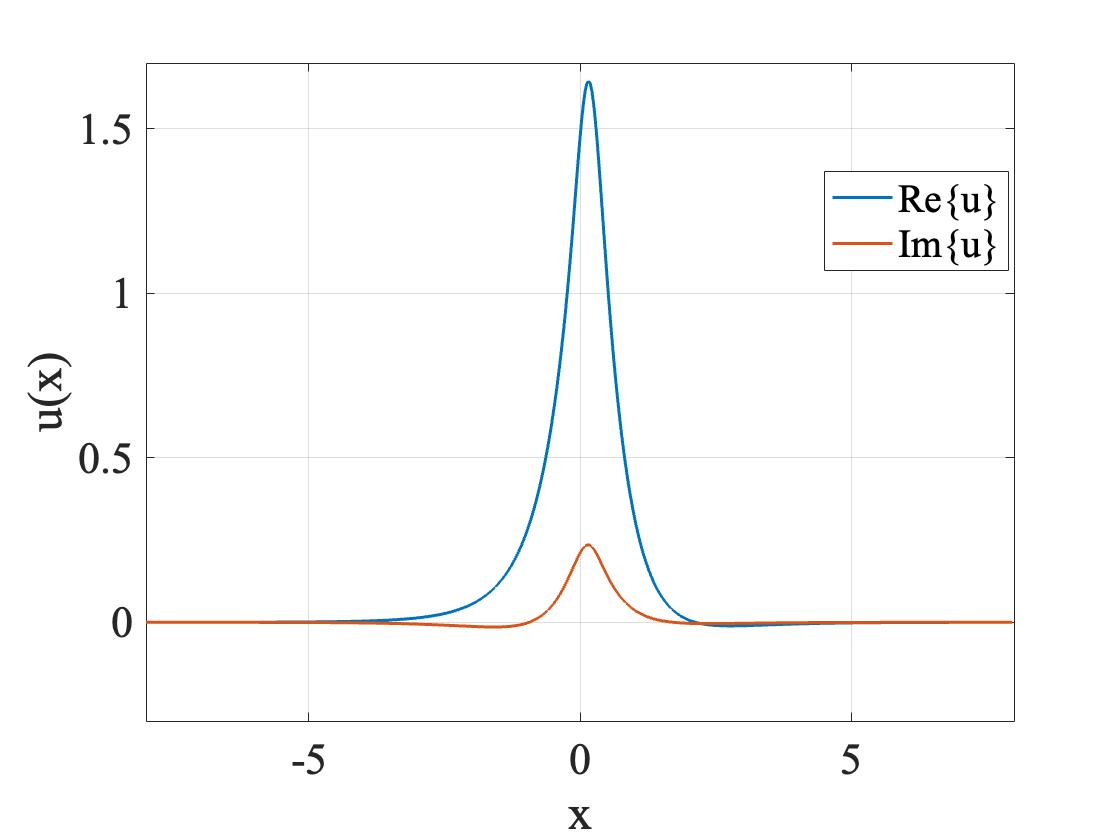}}
\subfigure[]{
        \includegraphics[width=3in]{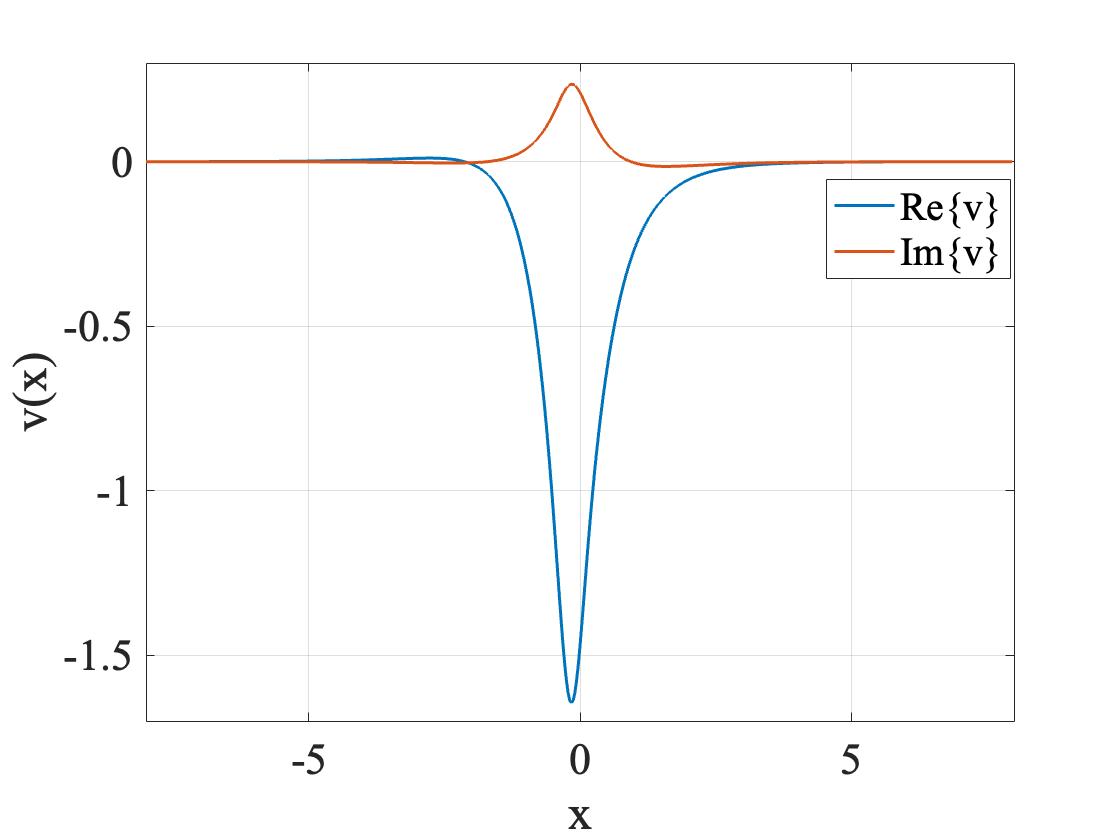}}
\subfigure[]{
        \includegraphics[width=3in]{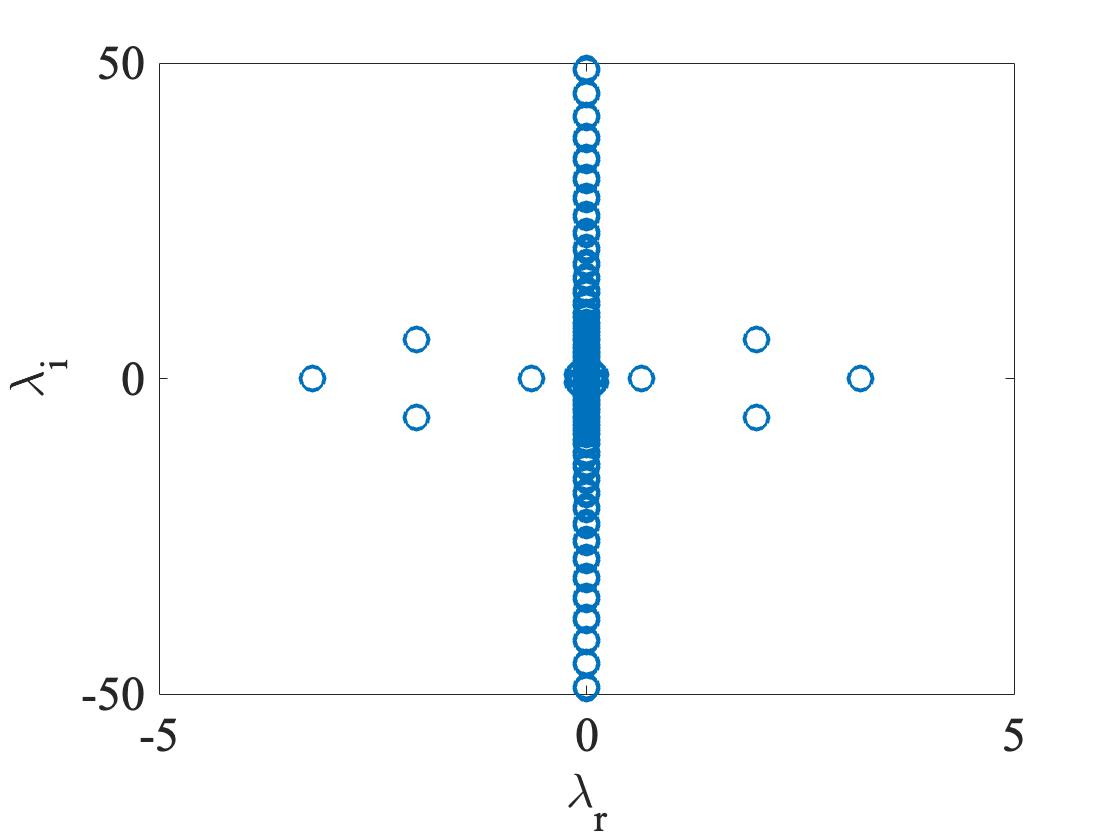}}
\subfigure[]{
        \includegraphics[width=3in]{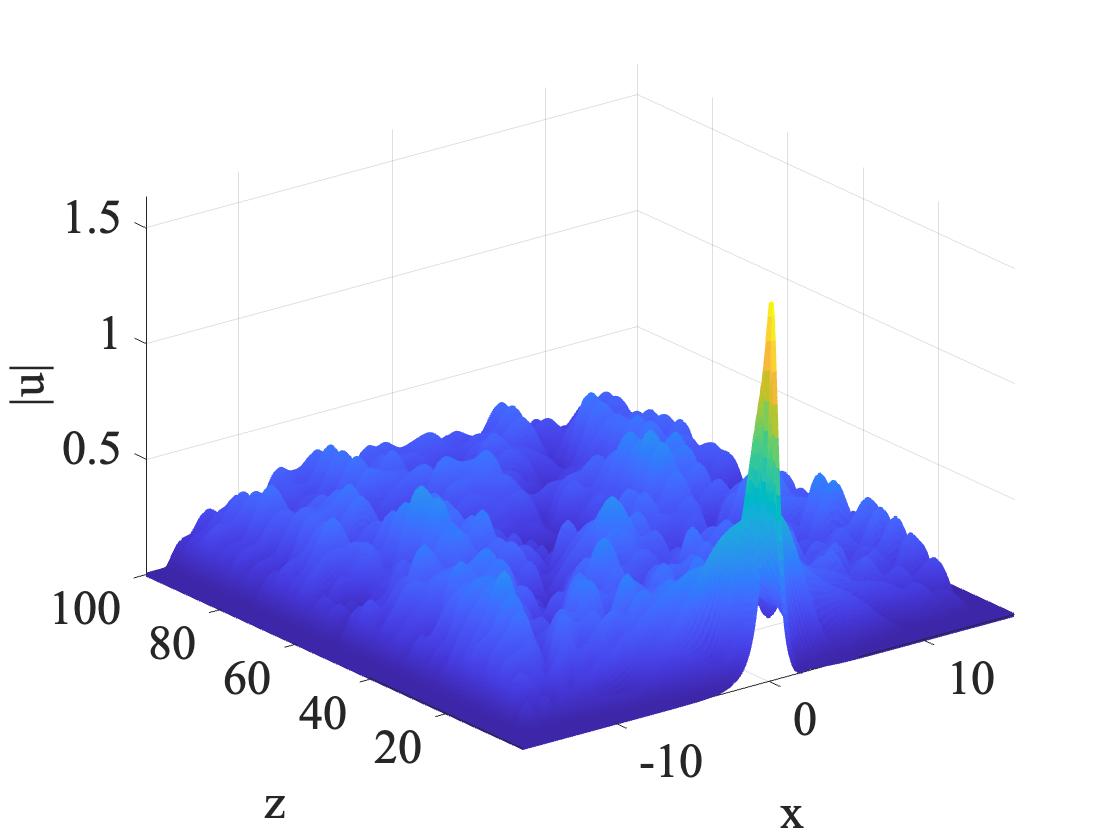}}}
\caption{(Color online). A typical example of an unstable $\mathcal{PT}$%
-antisymmetric soliton in the system with $\protect\delta =0.4,\protect%
\gamma =0.3$ for $k=1.5$. Components $u$ and $v$ of the stationary solution
are shown in panels (a) and (b), respectivly. (c) The spectrum of
eigenvalues for small perturbations, including unstable ones. (d) The
evolution of the soliton's $u$-component under the action of small random
perturbations at the amplitude level of $1\%$. The evolution of the $v$%
-component is similar (not shown here).}
\label{fig12}
\end{figure}

It is relevant to mention the system also support $\mathcal{PT}$%
-antisymmetric solitons, but they are all strongly unstable, similar to what
is known in many other $\mathcal{PT}$-symmetric systems \cite%
{Barash2,Barash3,Barash4,Barash5}. A typical example is displayed in Fig. %
\ref{fig12}. 

In addition to the symmetric and antisymmetric modes, the conservative system with $\gamma = 0$
admits asymmetric solutions \cite{EZB:2020}. However, in the case of $\gamma \neq 0$ asymmetric 
states does not exist, as they cannot maintain the balance between the gain and loss.

\begin{figure}[tbp]
\centering{\includegraphics[width=3.5in]{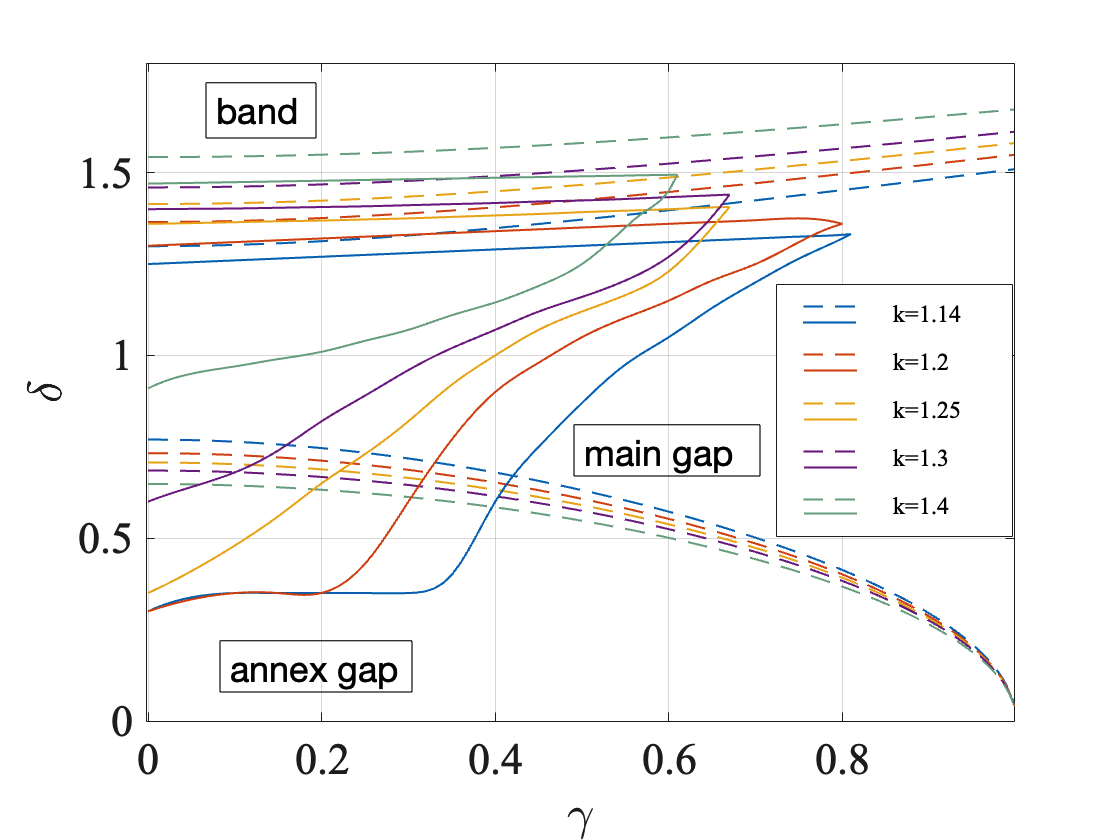}}
\caption{(Color online). Families of solitons with indicated values of
propagation constant $k$ are stable in areas surrounded by the respectively
colored solid curves, being unstable outside. Dashed curves of the same
colors in the upper part of the panel correspond to $\protect\gamma =\protect%
\gamma _{\mathrm{SIG}}$, see Eq. (\protect\ref{fixed_k_main}). They
designate boundaries between the main (semi-infinite) gap and the band of
linear waves, where solitons cannot exist. In the lower part, the dashed
curves, which correspond to $k=k_{\mathrm{SIG}}$ (see Eq. (\protect\ref%
{kmax1})), indicate boundaries between the main and annex gaps. }
\label{fig1}
\end{figure}

The results are summarized in Fig. \ref{fig1}, which displays numerically
found stability boundaries for soliton families in the plane of the
gain-loss and SOC coefficients, $\left( \gamma ,\delta \right) $, for
several fixed values of the propagation constant $k$. The figure also
includes the boundary between the SIG and band of linear waves, in which
solitons cannot exist. This boundary is determined by Eq. (\ref{kmax2}),
while the boundary between the SIG and AG is given by Eq. (\ref{fixed_k_main}%
).

The stability border was eventually identified according to results of
direct simulations of solitons with random noise at the amplitude level of $%
1\%$ added to the input. The simulations were run over the propagation
distance corresponding to $\geq 50$ diffraction (Rayleigh) lengths of the
unperturbed soliton. The computation of the instability growth rate from the
numerical solution of Eq. (\ref{BdG}) produced results compatible with those
obtained from the direct simulations (there may be a residual eigenvalue $%
\sim 0.01$ close to the boundary, which does not give rise to any
instability for long enough propagation distance in the simulations).

At $\gamma =0$, the stability boundaries shown in Fig. \ref{fig1} are
identical to those presented in Ref. \cite{EZB:2020}. Naturally, the
stability area shrinks with the increase of the gain-loss strength $\gamma $%
, and disappears at $\gamma \approx 0.8$, i.e., before reaching the
exceptional point, $\gamma =1$. Stability boundaries are not displayed for $%
k<1.14$, as the solitons become very broad for such values, and convergence
of the numerical iterations generating stationary solitons becomes very
slow. The smallest value of $k$ at which solitons were found is $k\approx
1.04$. Note that values of $k$ in the bandgaps, at which solitons may exist,
are bounded from below, as seen in Eq. (\ref{kmax2}).

\subsection{Dynamics of unstable solitons}

\begin{figure}[tbp]
\centering{\ \subfigure[]{\includegraphics[width=3in]{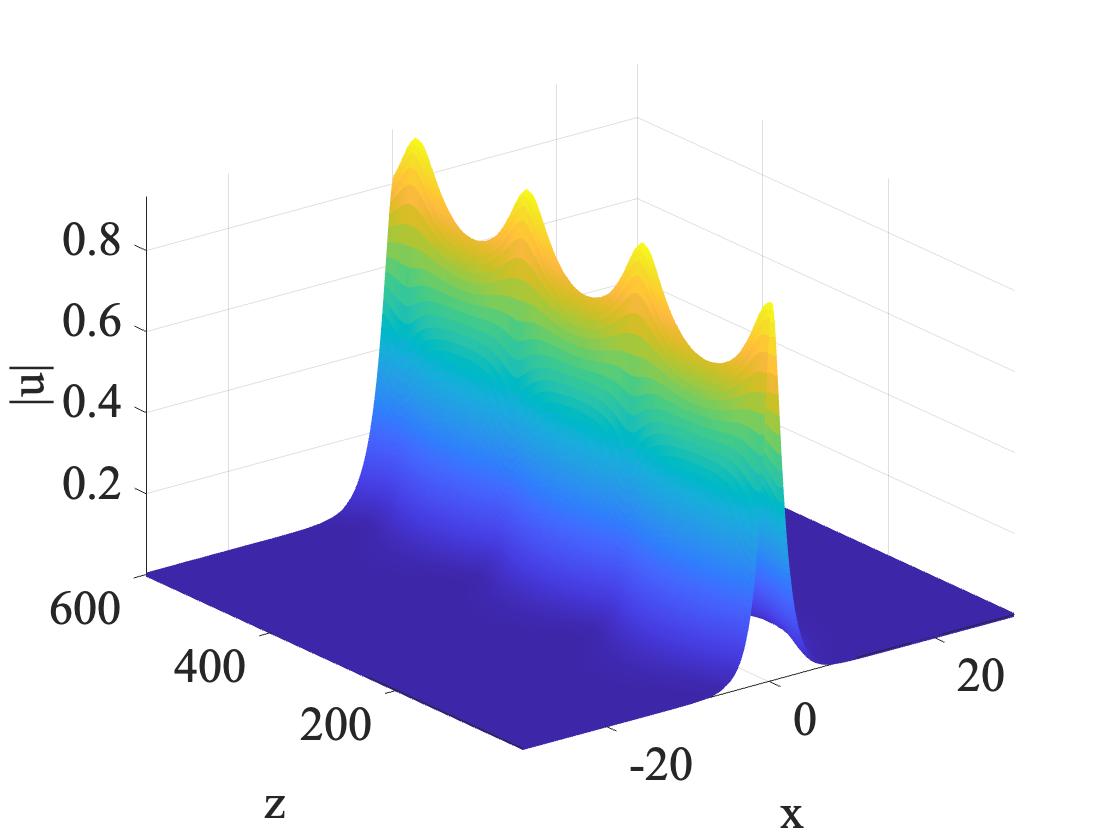}}
\subfigure[]{
        \includegraphics[width=3in]{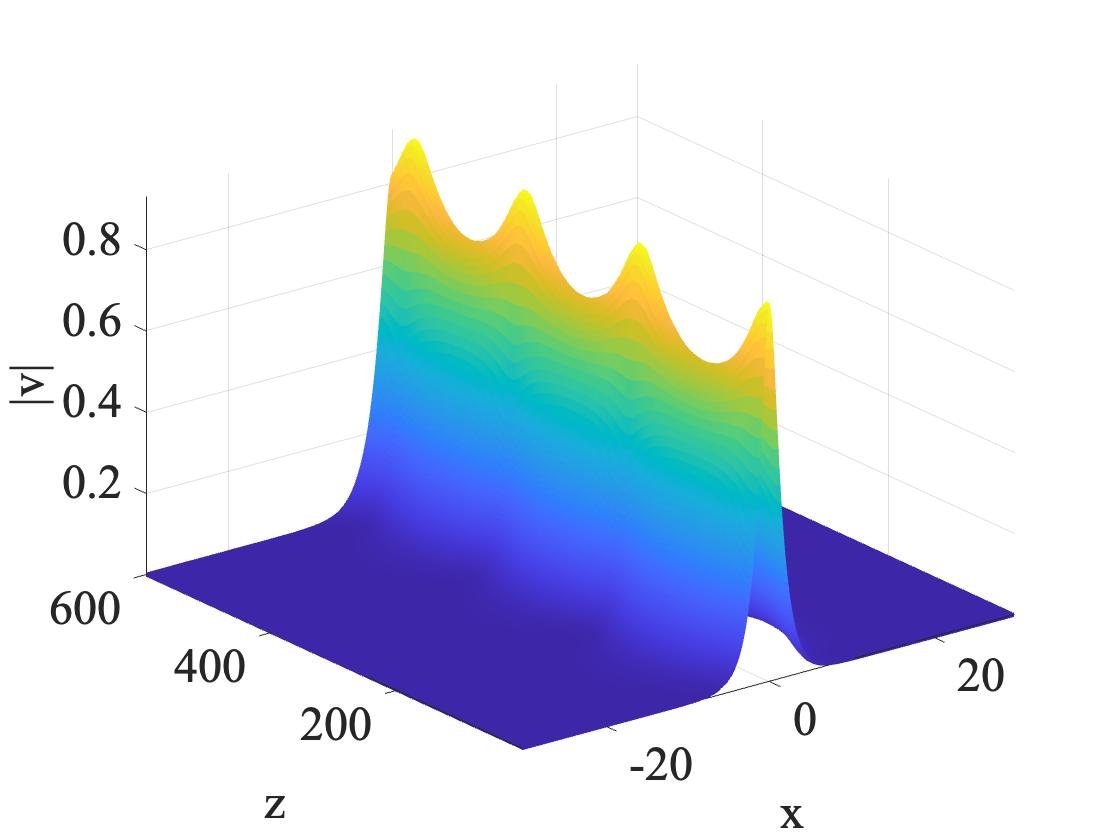}}
\subfigure[]{
        \includegraphics[width=3in]{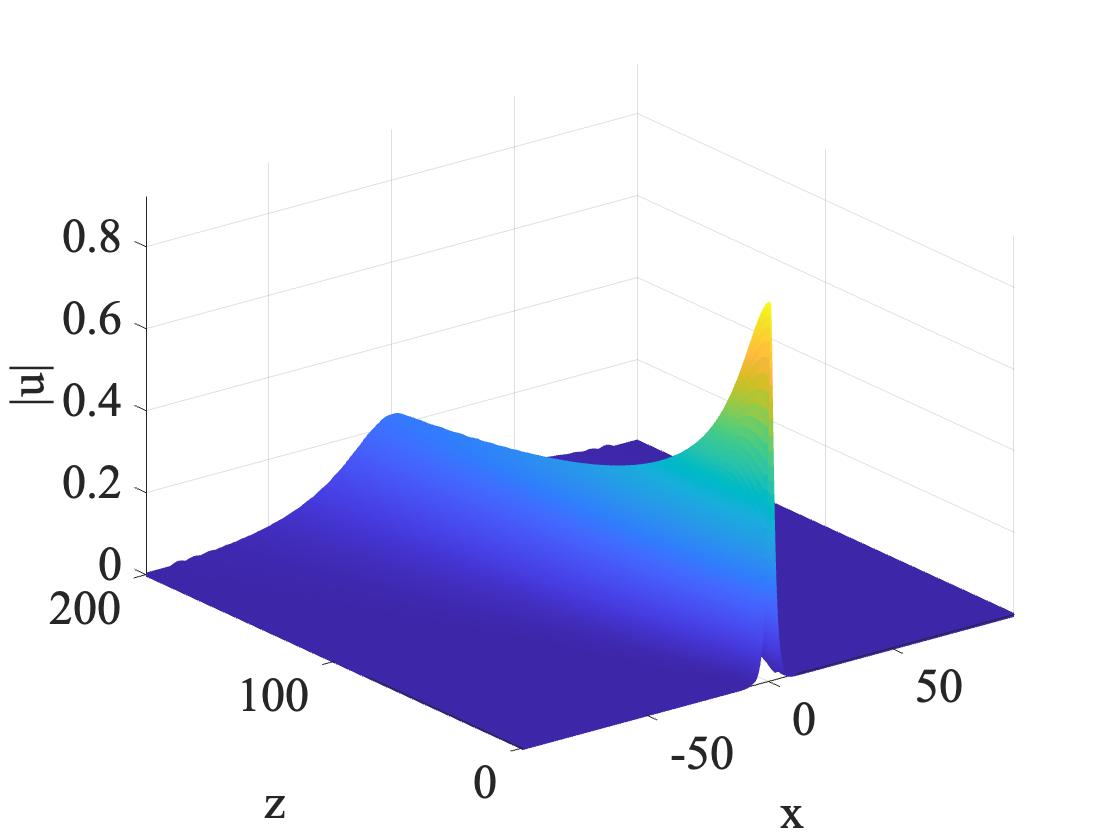}}
\subfigure[]{
        \includegraphics[width=3in]{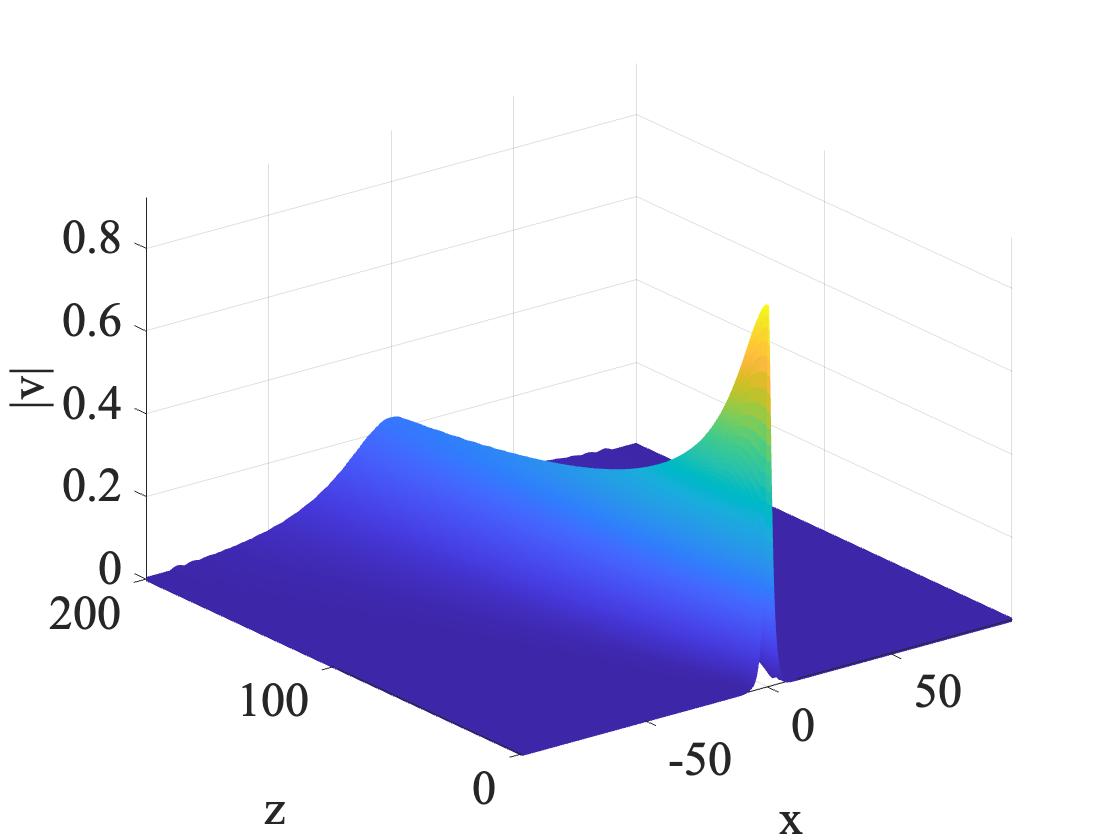}}}
\caption{(Color on line). Long-time evolution of a weakly unstable soliton
at $\protect\delta =0.35,\protect\gamma =0.1$, with propagation constant $%
k=1.25$. The simulations were initiated with random perturbations at the
amplitude level of $1\%$ added to the stationary soliton. Depending on the
particular realization of the random perturbation, three different outcomes
of the evolution are possible: transformation into a breather in (a,b);
decay in (c,d); and blowup (not shown here).}
\label{fig4}
\end{figure}

For unstable solitons, the simulations make it possible to distinguish
several dynamical scenarios. First, the collapse (blowup) is a generic
scenario in the entire instability area, see, e.g., Fig. \ref{fig3}.
Collapse may be avoided by unstable solitons residing close to the stability
boundary. Namely, at relatively small values of $\delta $ and $\gamma $,
weak instability spontaneously transforms stationary solitons with a
relatively small amplitude into breathers, as shown in Figs. \ref{fig4}%
(a,b). This outcome depends on the particular realization of small random
perturbations applied to the soliton. Another realization initiates,
instead, decay of the same soliton, as shown in Fig. \ref{fig4}(c,d). The
blowup of the same unstable soliton is possible too, under the action of a
different realization (not shown here). This peculiarity is possible because
an unstable soliton may give rise to several different unstable eigenvalues,
associated with different eignemodes of small perturbations (see, e.g., Fig. %
\ref{fig3}(b)). Accordingly, a specially crafted small initial perturbation
may excite a specific eigenmode.

\begin{figure}[tbp]
\centering{\ \subfigure[]{\includegraphics[width=3in]{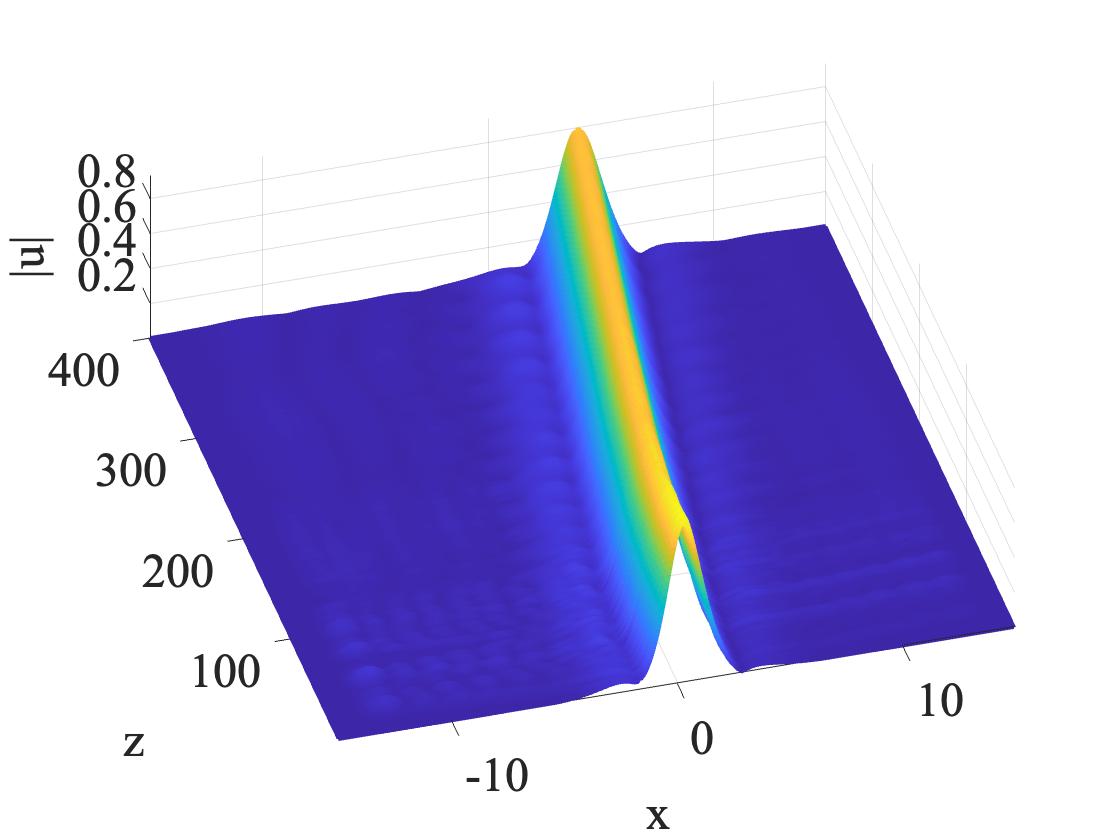}}
\subfigure[]{
        \includegraphics[width=3in]{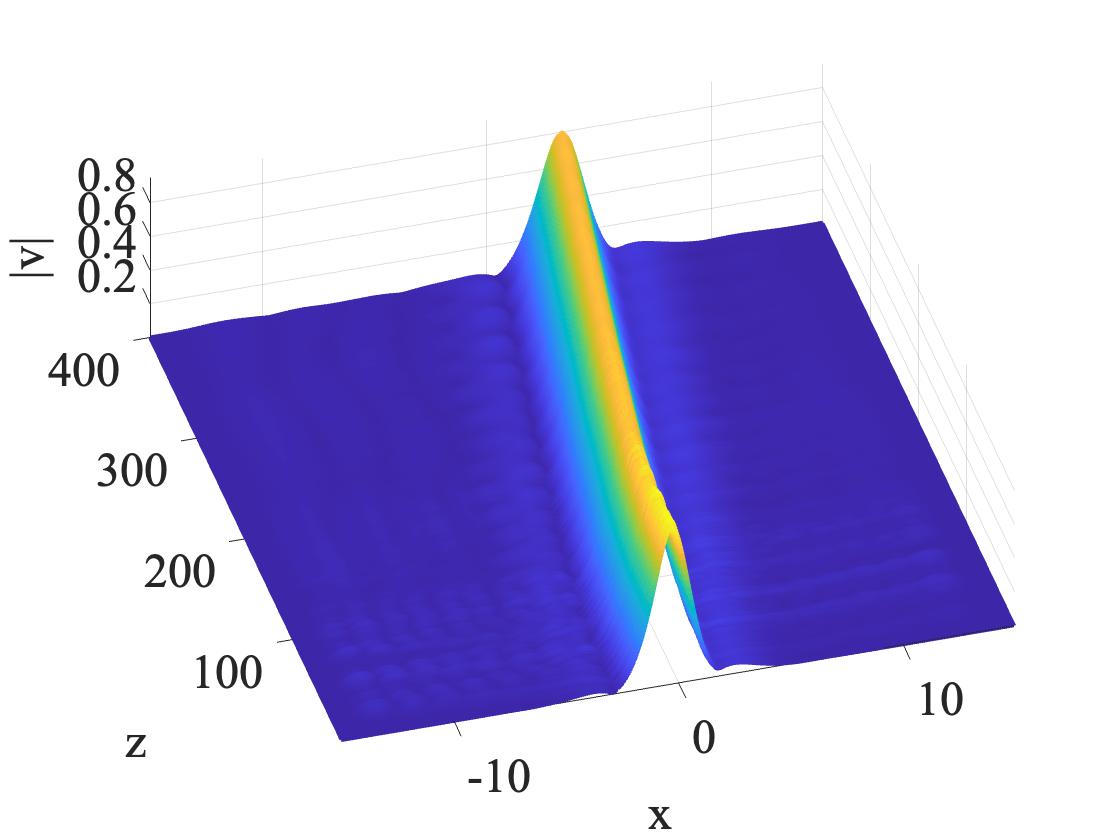}} 
}
\caption{(Color on line). The evolution of a weakly unstable soliton at $%
\protect\delta =1,\protect\gamma =0.6$, with propagation constant $k=1.14$.
The simulations were initiated with random perturbations at the amplitude
level of $1\%$ added to the stationary soliton, which is located close to
the stability boundary for $k=1.14$, see Fig. \protect\ref{fig6}. The
soliton spontaneously develops tilt in the spatial domain (\textquotedblleft
motion"). The change of the sign of $\protect\gamma $ leads to a similar
result with the opposite sign of the tilt.}
\label{fig5}
\end{figure}

At larger values of $\gamma $ and $\delta $, it may also happen that an
unstable soliton is not destroyed. Instead, as shown in Figs. \ref{fig5}%
(c,d), the perturbed soliton starts spontaneous motion with weak vibrations
(actually, it develops a tilt in the spatial domain). The size of the tilt
(\textquotedblleft velocity") depends on both a particular realization of
the small random perturbation applied to the soliton and the system's
parameters. The inversion of the sign of the $\mathcal{PT}$ coefficient, $%
\gamma \rightarrow -\gamma $, produces the same result, but with the
opposite sign of the tilt. This observation is explained by the ($\mathcal{PT%
}$) invariance of the underlying equations (\ref{udelta}) and (\ref{vdelta})
with respect to substitution%
\begin{equation}
\left( u,v\right) \rightarrow \left( u^{\ast },v^{\ast }\right)
,z\rightarrow -z,\gamma \rightarrow -\gamma ,  \label{invar}
\end{equation}%
if it is applied to Eq. (\ref{dM/dz}).

\begin{figure}[tbp]
\centering{\ \subfigure[]{\includegraphics[width=3in]{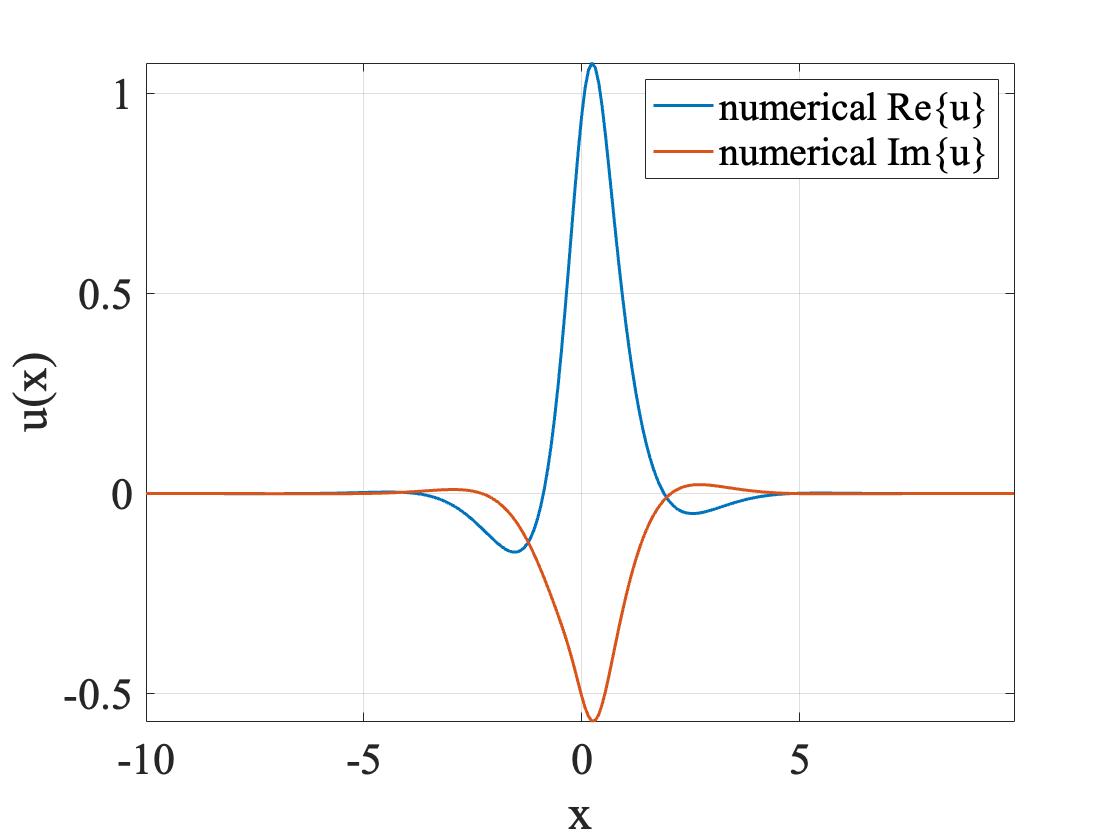}}
\subfigure[]{
        \includegraphics[width=3in]{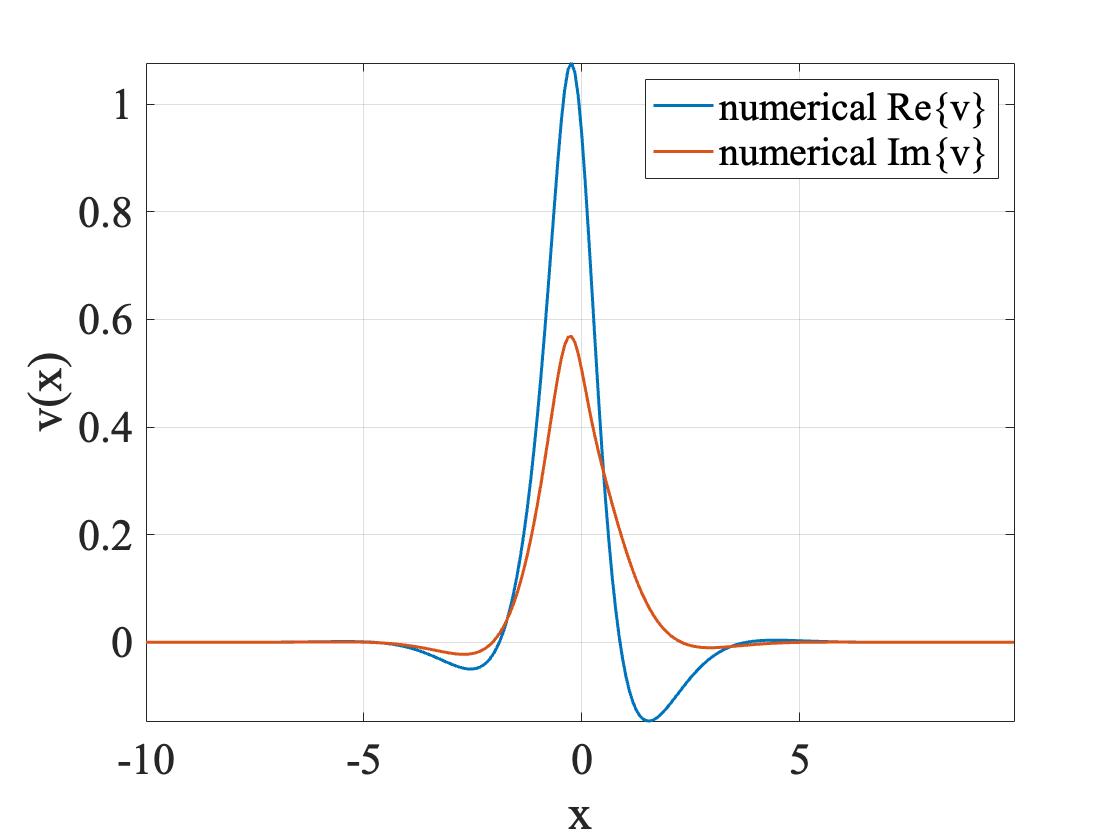}}
\subfigure[]{
        \includegraphics[width=3in]{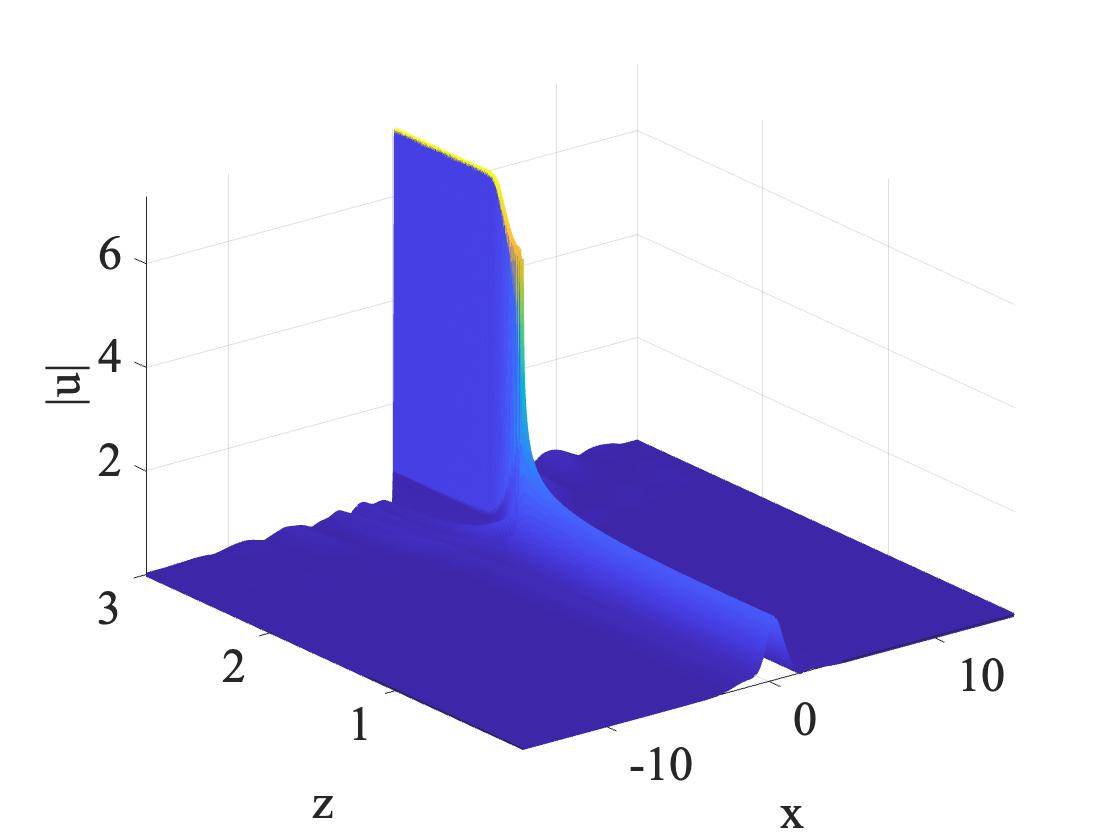}}
\subfigure[]{
        \includegraphics[width=3in]{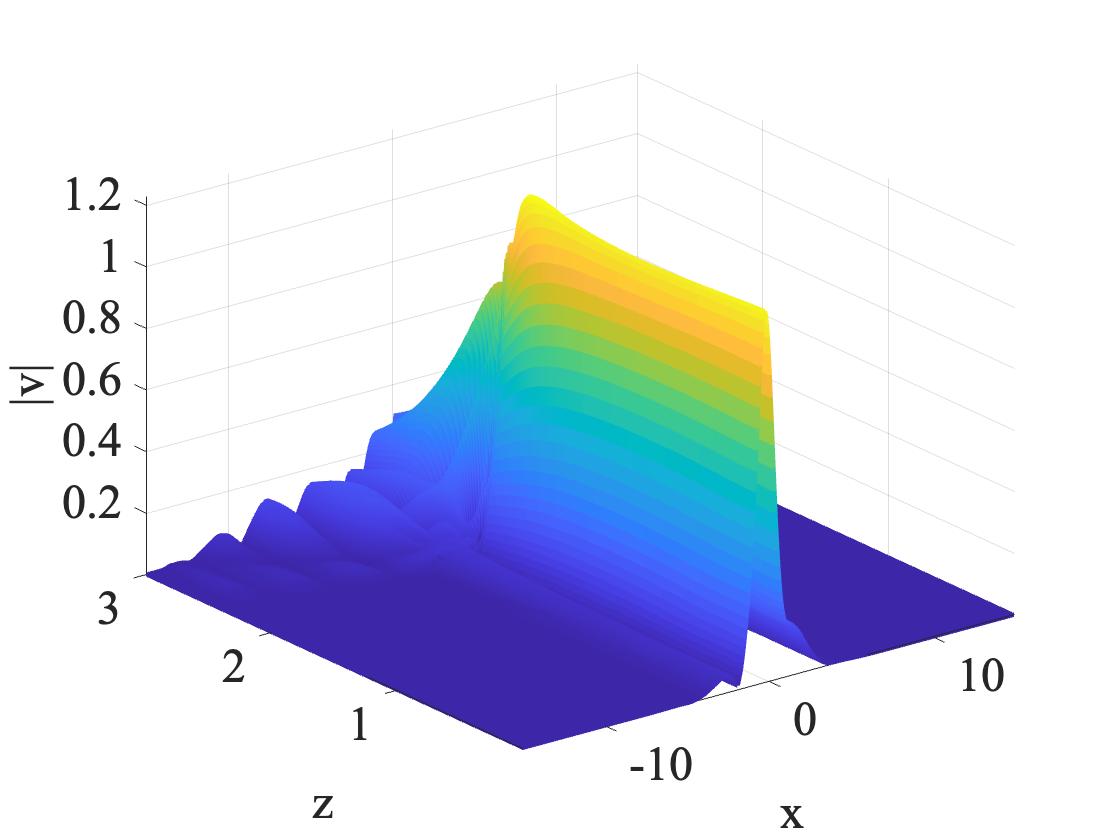}}}
\caption{(Color online). (a,b): An example of an unstable soliton found, as
a numerical solution of Eqs. (\protect\ref{Udelta}) and (\protect\ref{Vdelta}%
), at $\protect\gamma =1.2$, i.e., beyond the exceptional point ($\protect%
\gamma =1$). Other parameters of the soliton are $\protect\delta =1$ and $%
k=1.2$. Panels (c) and (d) demonstrate instability of this state against
spontaneous onset of the blowup.}
\label{fig10}
\end{figure}

Lastly, it is relevant to mention that, under the action of the
above-mentioned regularizing effect provided by SOC, Eqs. (\ref{Udelta}) and
(\ref{Vdelta}) produce stationary solitons solutions at $\gamma >1$, while
it is usually assumed that solitons cannot exist beyond the exceptional
point (in particular, the exact solution given by Eqs. (\ref{U0})-(\ref{00})
for $\delta =0$ does not exist at $\gamma >1$). Indeed, Eq. (\ref{kmax1})
defines a formal bandgap also for $\gamma >1$, although a part of the
spectrum is complex in this case, see Eq. (\ref{complex}). An example of the
soliton found at $\gamma =1.2$ is displayed in Figs. \ref{fig10}(a,b). As
expected, it is unstable against spontaneous onset of the blowup, see Figs. %
\ref{fig10}(c,d). Nevertheless, it is relevant to stress that this is a
genuine solution of Eqs. (\ref{Udelta}) and (\ref{Vdelta}), rather than the
so-called ``ghost state", which may be found as a formal solution beyond the
exceptional point, but does not represent a true stationary state of the
system \cite{ghost,ghost2}.

\subsection{Interactions between solitons}

\begin{figure}[tbp]
\centering{\ \subfigure[]{\includegraphics[width=3in]{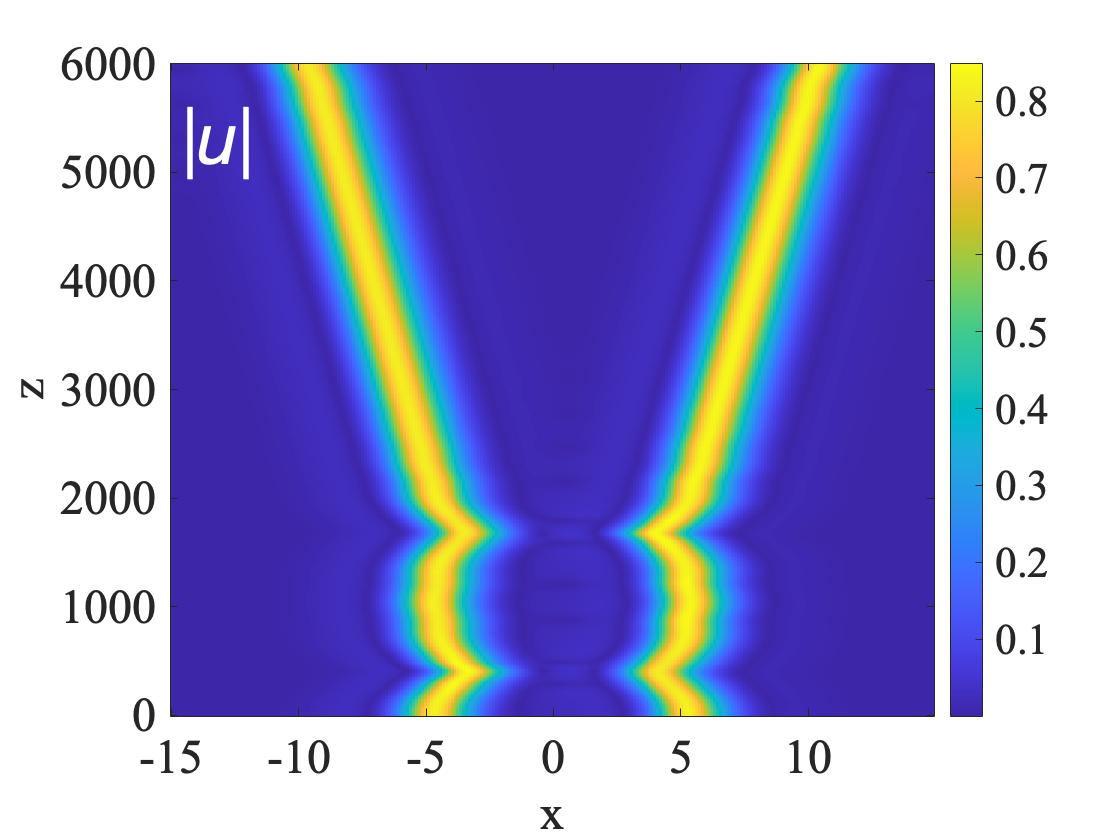}}
\subfigure[]{
        \includegraphics[width=3in]{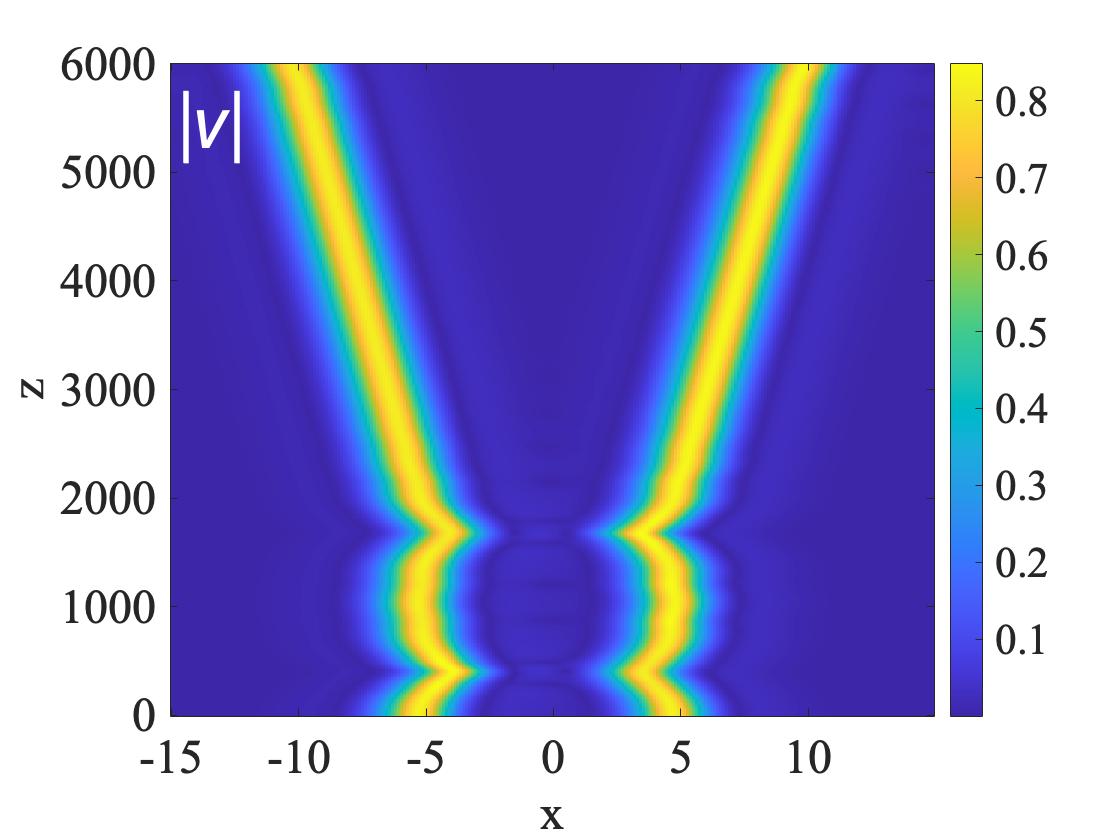}}
\subfigure[]{
        \includegraphics[width=3in]{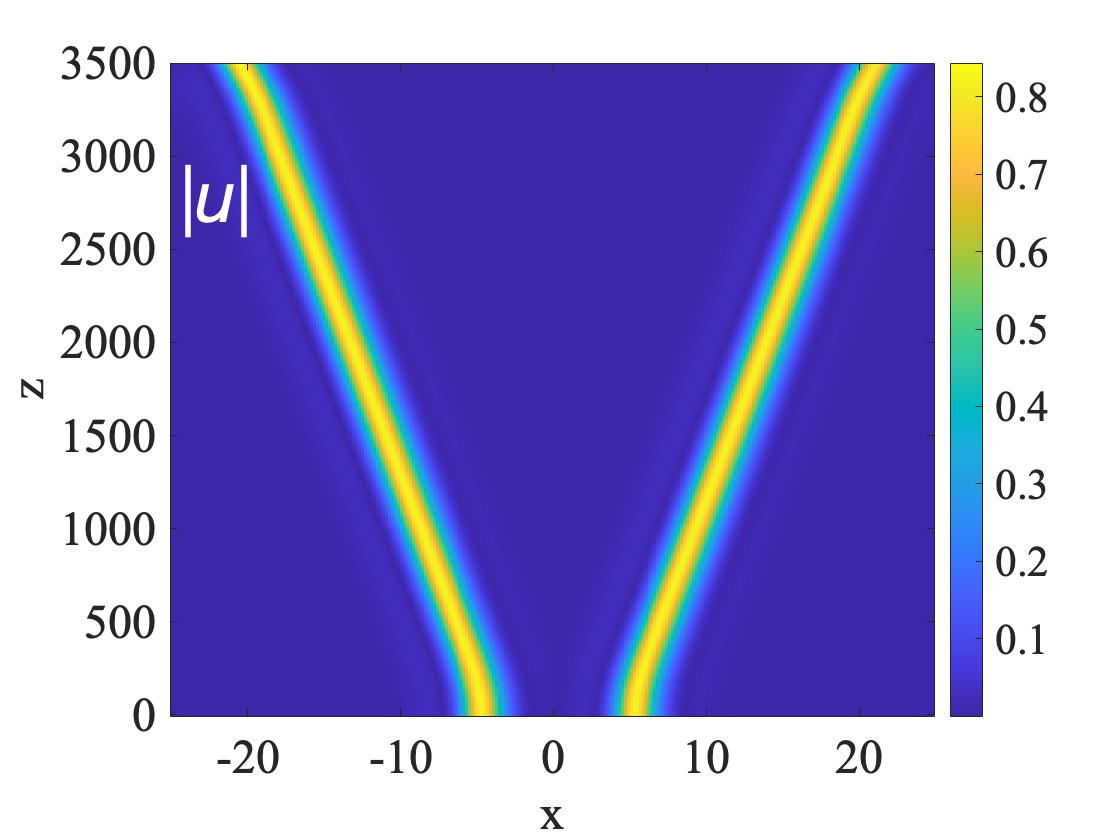}}
\subfigure[]{
        \includegraphics[width=3in]{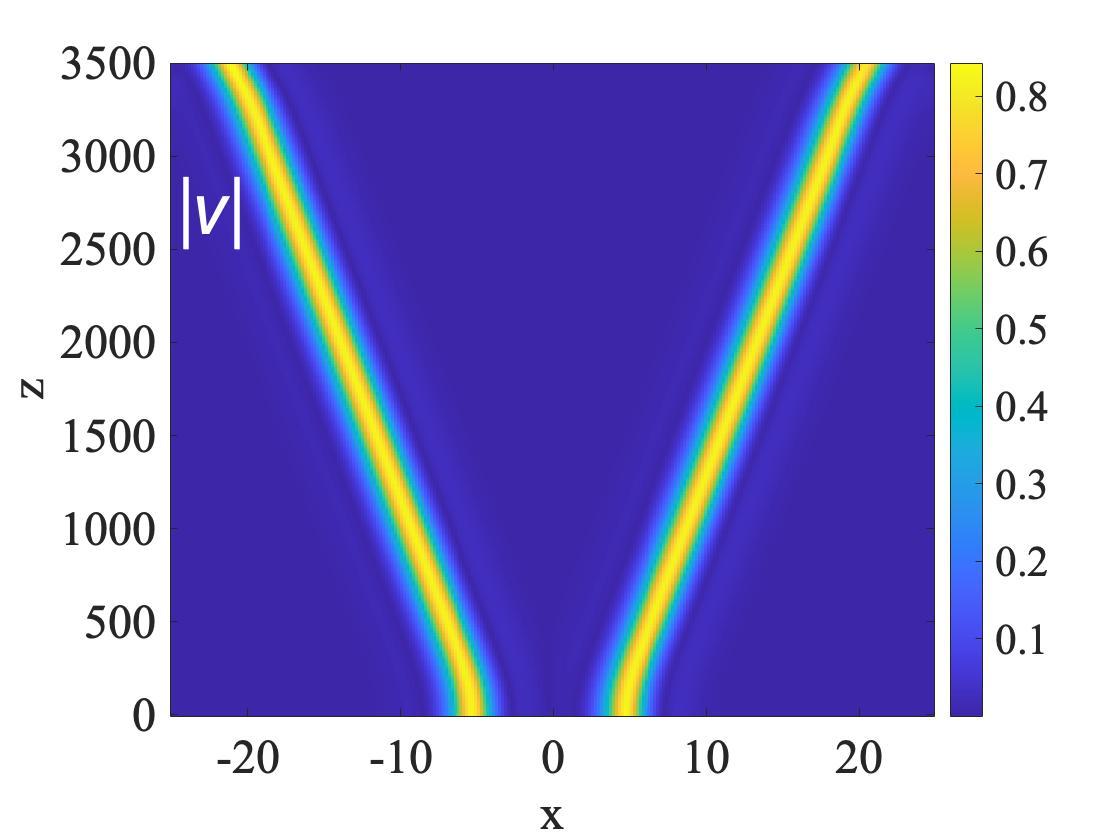}}}
\caption{(Color on line). Simulations of the interaction of two stable
solitons with parameters $\protect\delta =1,\protect\gamma =0.1$ and $k=1.2$%
, initially separated by distance $d=10$. (a,b): Relatively complex
interaction of the in-phase solitons, with the initial phase shift $\Delta
\protect\phi =0$. (c,d): Straightforward repulsion between the out-of-phase
solitons, with $\Delta \protect\phi =\protect\pi $.}
\label{fig9}
\end{figure}

The availability of stable solitons suggests a possibility to explore
interactions between them, initially placing two solitons at some distance $%
d $ between their centers (this possibility was not explored in work \cite%
{EZB:2020}, which introduced the conservative system with $\gamma =0$). It
is well known that, in usual conservative models, pairs of well separated
in-phase and out-of-phase solitons (ones with phase difference $\Delta \phi
=0$ or $\pi $) feature, respectively, mutual attraction or repulsion \cite%
{KM}. In our system, a qualitatively similar situation is exhibited by Fig. %
\ref{fig9}, where $d=10$, while the FWHM width of each soliton is $\simeq
1.5 $. In Figs. \ref{fig9}(a,b), two in-phase solitons originally attract
each other, then bounce back twice, still keeping a considerable distance,
and eventually separate. The essentially complex intrinsic structure of
solitons governed by Eqs. (\ref{udelta}) and (\ref{vdelta}) leads to the
change of the initial phase shift between the interacting solitons, leading
to the fact that eventually separating solitons develop a phase shift of $%
\pi $. In Figs. \ref{fig9}(c,d), the simulation demonstrates straightforward
repulsion and separation of the same pair of the stable solitons, but with
the initial phase shift $\Delta \phi =\pi $.

\begin{figure}[tbp]
\centering{\ \subfigure[]{\includegraphics[width=3in]{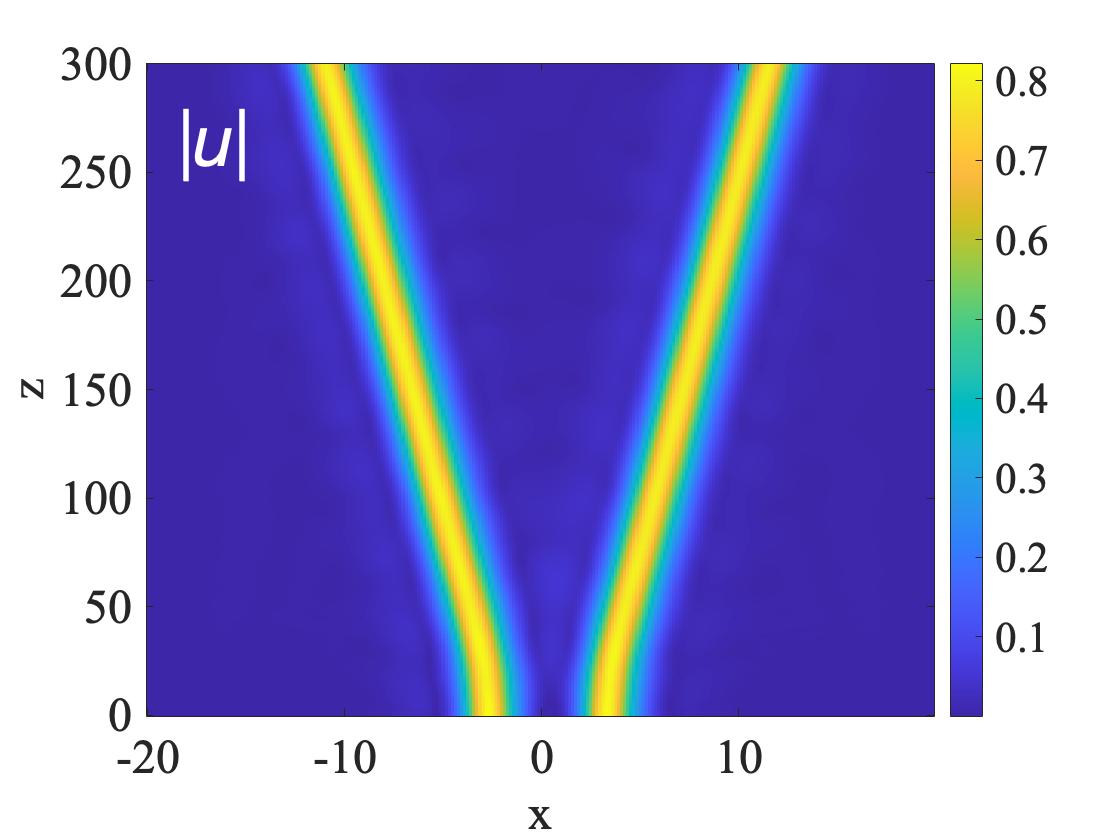}}
\subfigure[]{
        \includegraphics[width=3in]{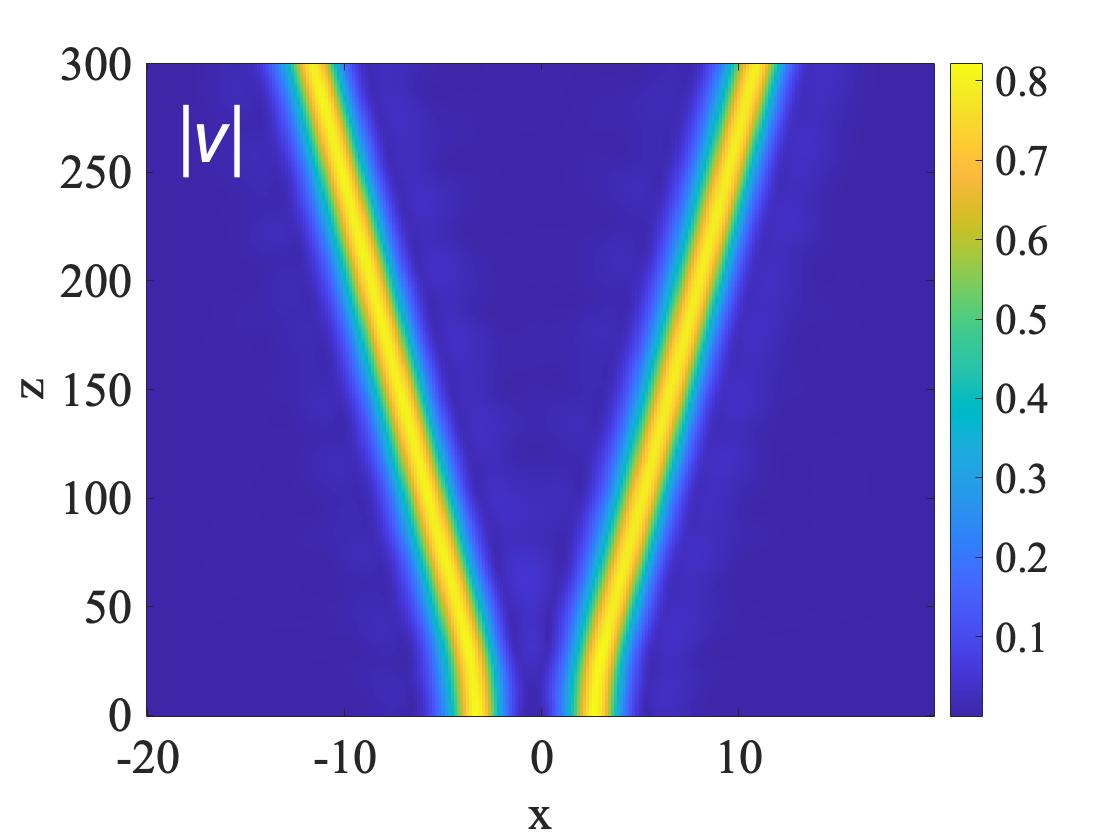}}
\subfigure[]{
        \includegraphics[width=3in]{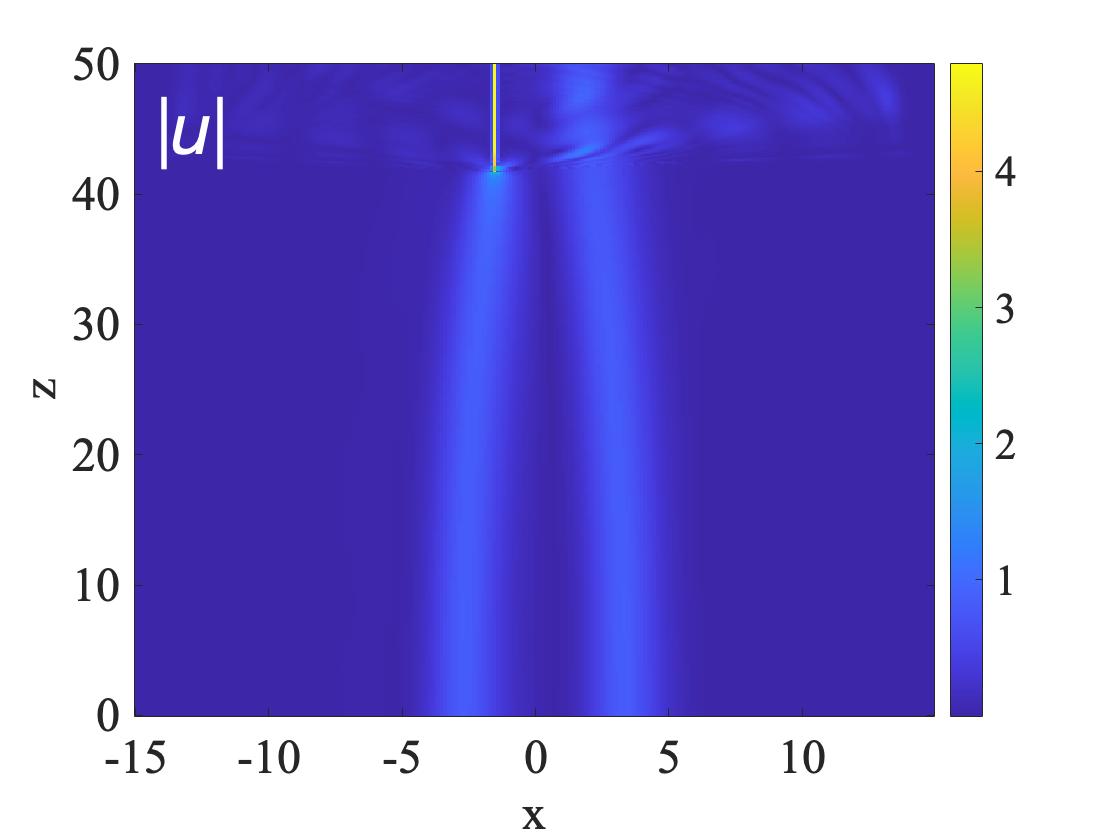}}
\subfigure[]{
        \includegraphics[width=3in]{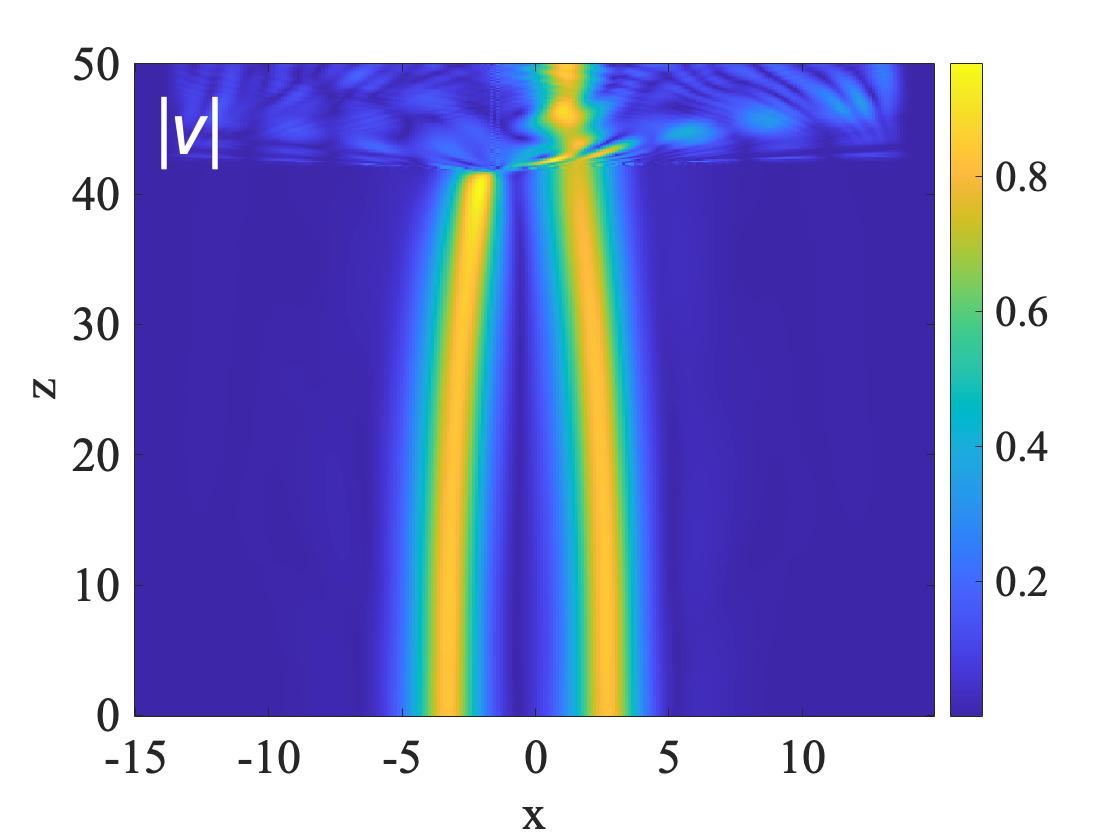}}}
\caption{(Color on line). The evolution of the same pairs of solitons as in
Fig. \protect\ref{fig9}, but with a smaller initial separation, $d=6$. In
panels (a) and (b), the in-phase solitons, counter-intuitively, repel each
other. In panels (c) and (d), out-of-phase solitons unexpectedly exhibit
attraction, and eventually merge into an collapsing state.}
\label{fig8}
\end{figure}

The interaction is more complex for the same pair of solitons with a smaller
initial separation, such as $d=6$ in Fig. \ref{fig8}. In this case, the
original overlap between the solitons is conspicuous, which does not allow
to consider them as a usual separated pair. As a result, the character of
the interaction is drastically different from the usual pattern: as shown in
Figs. \ref{fig8}(a,b), the in-phase solitons immediately \textit{repel} each
other, while the out-of-phase solitons feature \textit{attraction}. In the
latter case, they eventually merge into a single object, which is then
destroyed by the blow-up (collapse).

\subsection{Tilted (``moving") solitons}

\begin{figure}[tbp]
\centering{\ \subfigure[]{\includegraphics[width=3in]{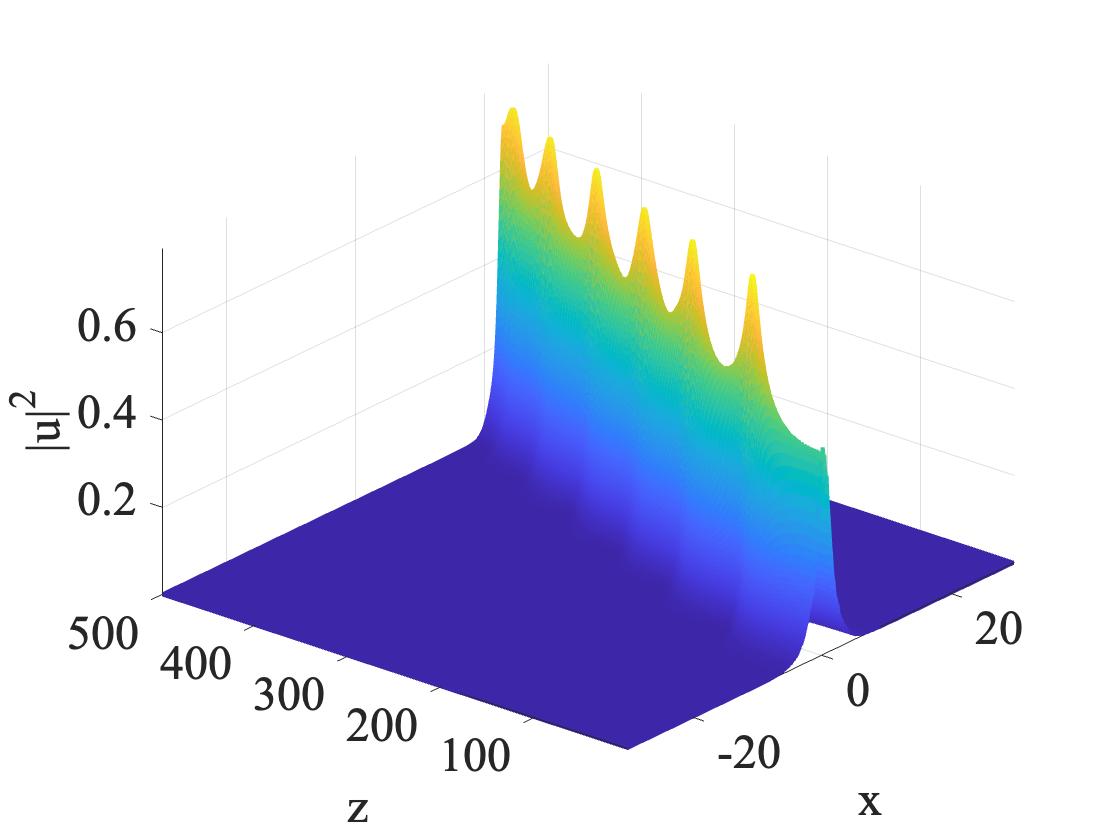}}
\subfigure[]{
        \includegraphics[width=3in]{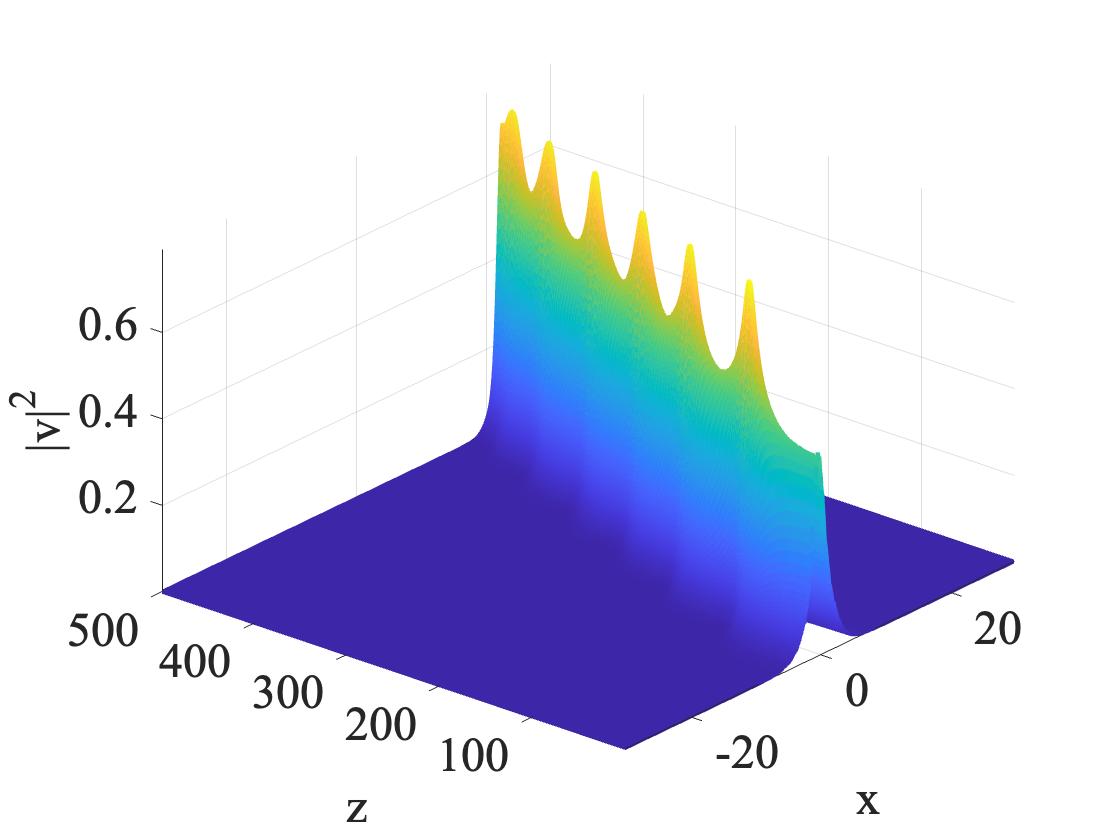}}
\subfigure[]{
        \includegraphics[width=3in]{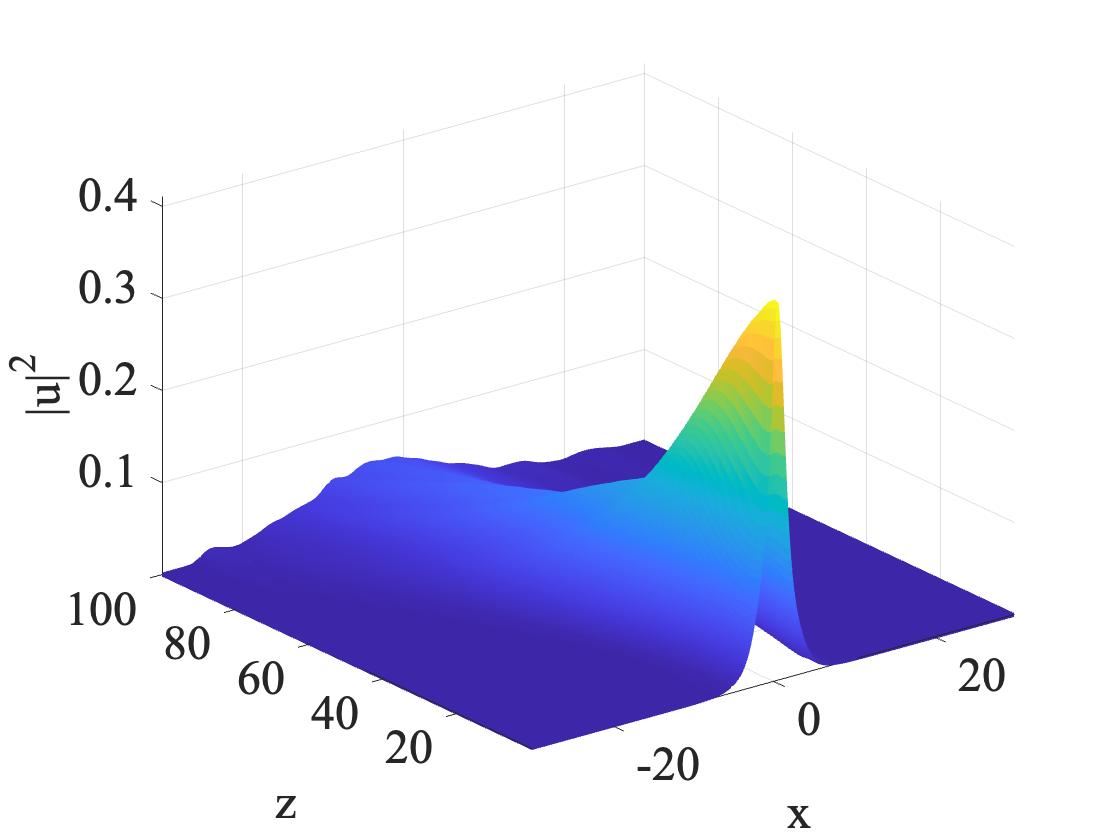}}
\subfigure[]{
        \includegraphics[width=3in]{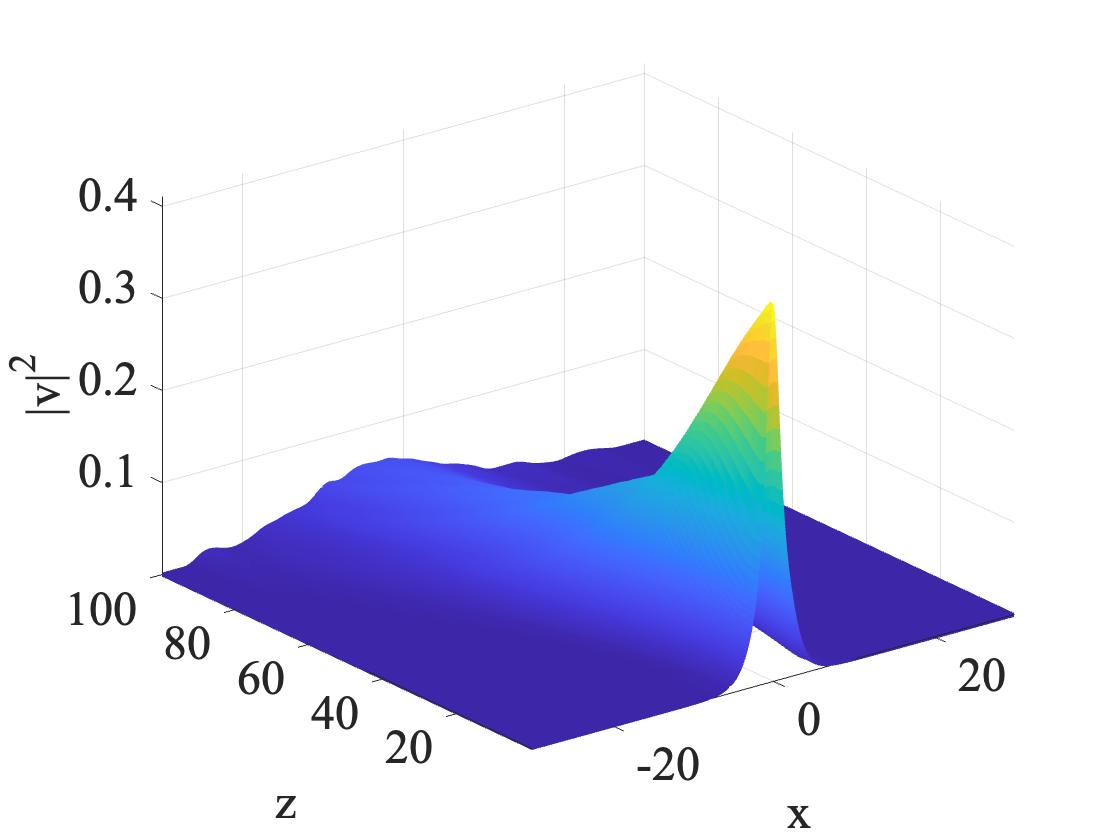}}}
\caption{(Color on line). The perturbed evolution of unstable tilted
solitons with \textquotedblleft velocity" $c=0.2$ (see Eq. (\protect\ref{xi}%
)) and propagation constant $k=1.3$, in the system with $\protect\delta =1.2$
and $\protect\gamma =0.2$. In panels (a) and (b), the soliton transforms
itself into an enhanced (relatively tall) robust breather. In (c) and (d),
an \textquotedblleft unfavorable" realization of the random perturbation
initiates decay of the soliton.}
\label{fig11}
\end{figure}

The study of tilted (moving) solitons in the presence of SOC is a nontrivial
issue, as SOC (unlike the gain and loss terms which represent the $\mathcal{%
PT}$ symmetry) destroys the system's Galilean invariance, making it
impossible to construct moving solitons as boosted copies of quiescent ones.
In Ref. \cite{EZB:2020}, the analysis of the solutions in the tilted
(\textquotedblleft moving") reference frame, with $x$ replaced by the tilted
coordinate (\ref{xi}), had demonstrated that all tilted solitons in the
conservative system with $\gamma =0$ are unstable. In most cases, it is weak
instability, which spontaneously transforms the tilted solitons into
breathers. In the present system, which includes the gain-and-loss terms,
systematic simulations also produce unstable tilted solitons. At small
values of the tilt, such as $c=0.2$ in Fig. \ref{fig11}, the instability is
relatively weak. With \textquotedblleft favorable" realizations of random
perturbations, a tilted soliton is spontaneously converted into a robust
tilted breather, as shown in Figs. \ref{fig11}(a,b). Note that the
established amplitude of the breather is essentially higher than that of the
input soliton, which implies that the formation of such an \textquotedblleft
enhanced" breather may be considered as an \textquotedblleft arrested
collapse". On the other hand, similar to what is shown above for quiescent
unstable solitons in Fig. \ref{fig4}, \textquotedblleft unfavorable"
perturbations may initiate decay of the same soliton, as is seen in Figs. %
\ref{fig11}(c,d). At large $c$, unstable solitons always suffer destruction
(not shown here in detail).

\subsection{Unstable solitons in the reduced system}

Similar to the conservative system ($\gamma =0$) considered in Ref. \cite%
{EZB:2020} (see also the exact solution (\ref{exact})), the simplified
system of Eqs. (\ref{u2}) and (\ref{v2}), which neglects the paraxial
diffraction, can readily produce soliton solutions in the finite bandgap (%
\ref{gap-c}), but they all turn out to be unstable, on the contrary to the
above results (families of stable solitons in the SIG and AG) produced for
the full system of Eqs. (\ref{udelta}) and (\ref{vdelta}). A typical example
of an unstable soliton generated by the reduced system, along with the
spectrum of its (in)stability eigenvalues and results of the perturbed
evolution, which shows destruction of the soliton, is displayed in Fig. \ref%
{fig7}.

\begin{figure}[tbp]
\centering{\ \subfigure[]{\includegraphics[width=3in]{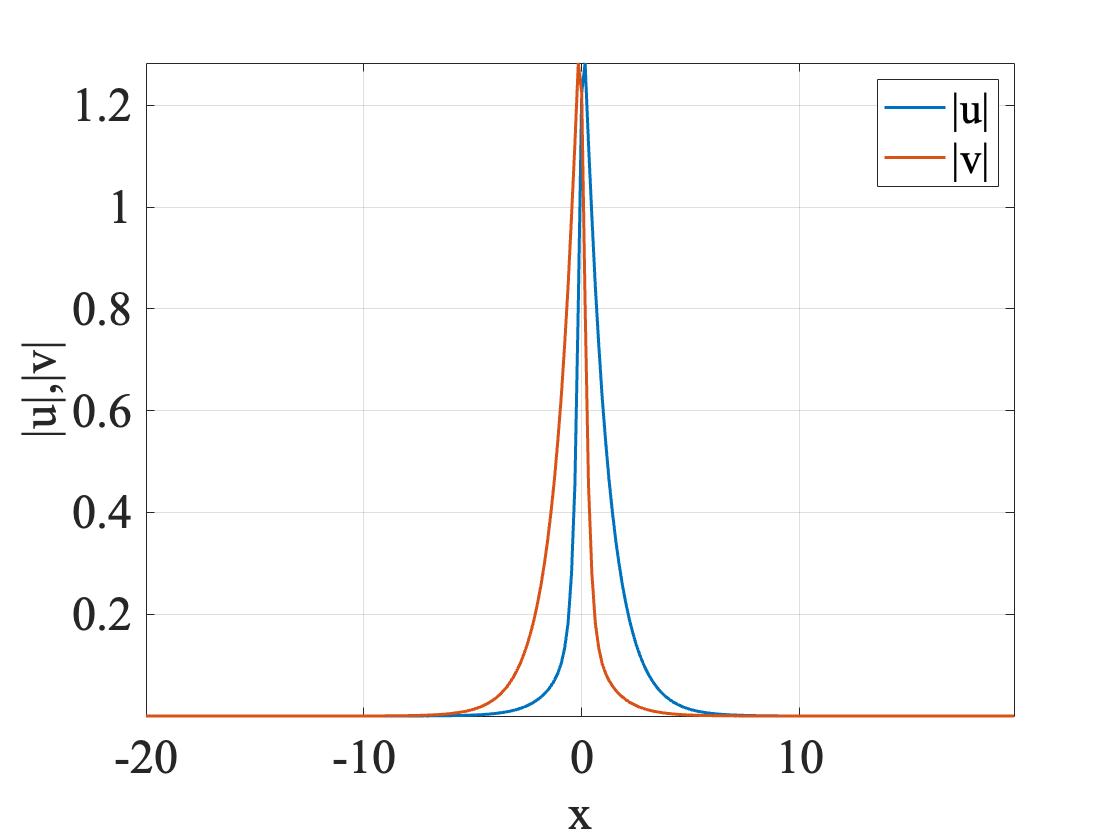}}
\subfigure[]{
        \includegraphics[width=3in]{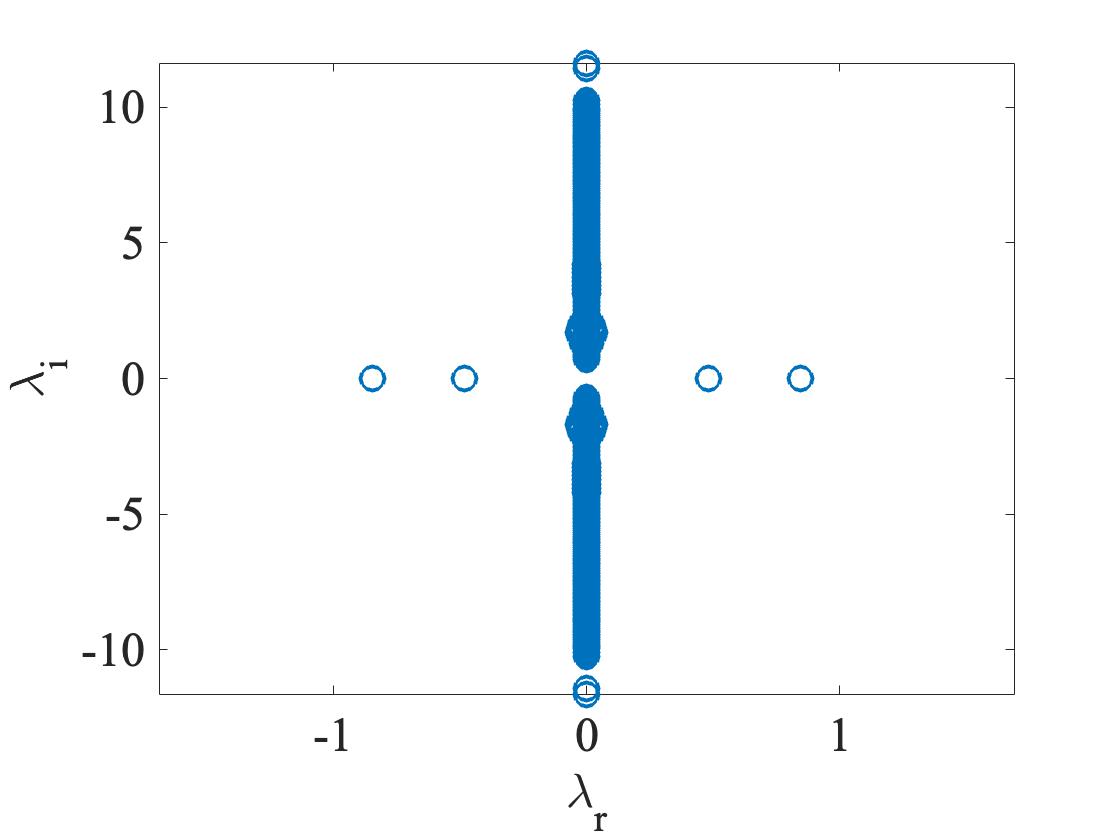}}
\subfigure[]{
        \includegraphics[width=3in]{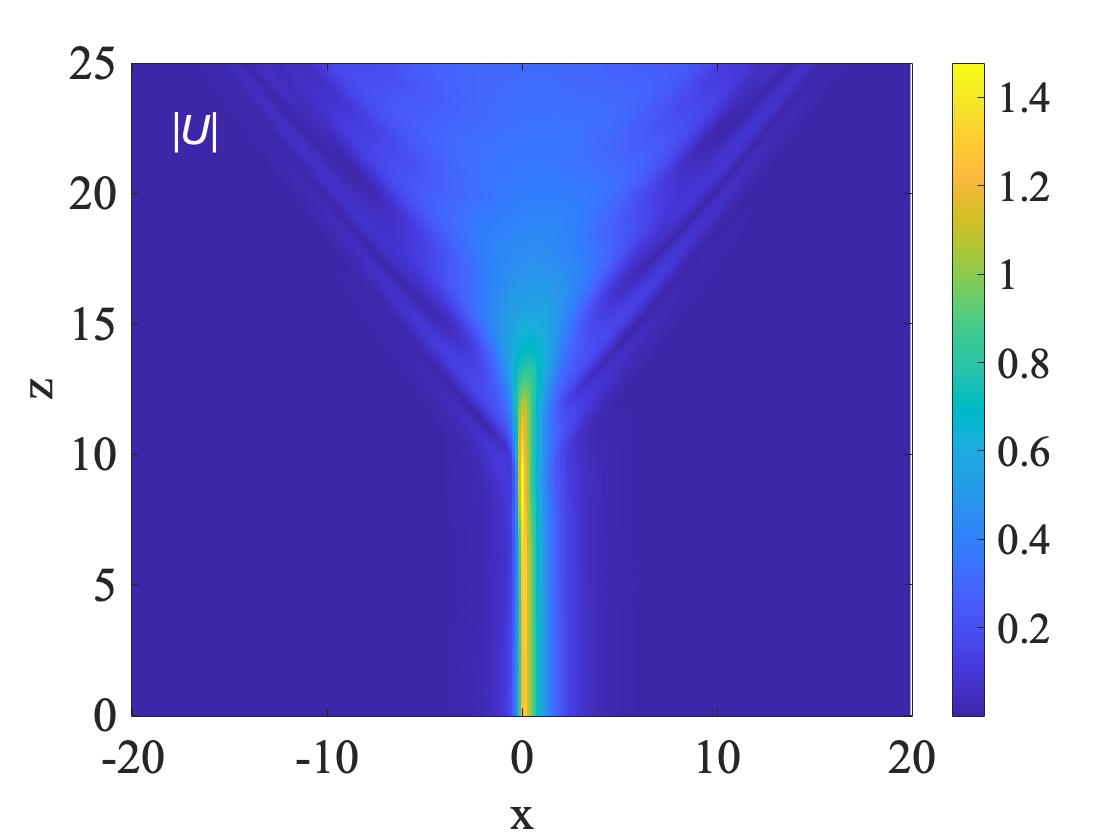}}
\subfigure[]{
        \includegraphics[width=3in]{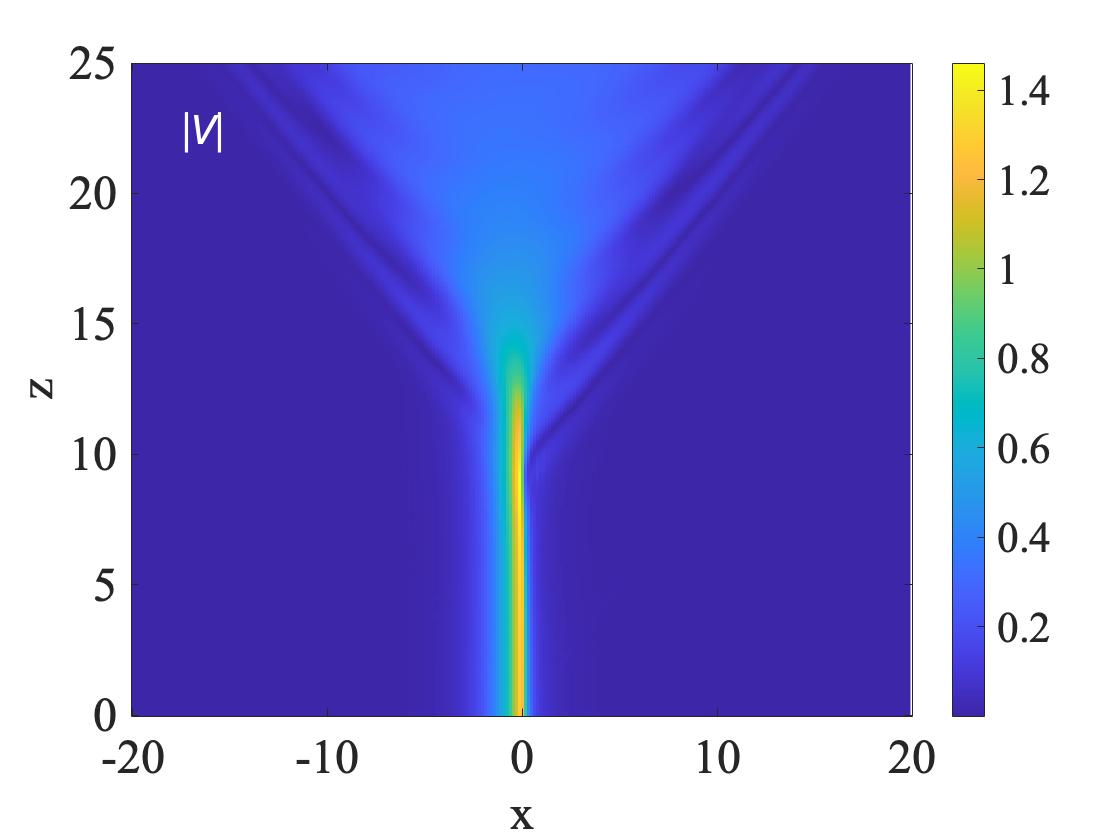}} }
\caption{(Color on line). A typical example of an unstable soliton produced
by the reduced system of Eqs. (\protect\ref{u2}), (\protect\ref{v2}) with $%
\protect\gamma =0.2$. The soliton's propagation constant is $k=-0.2$. (a)
The structure of the soliton; (b) the eigenvalue spectrum, which includes
two pairs of unstable modes of small perturbations; (c,d) the evolution of
the soliton under the action of random perturbations initially added at the $%
1\%$ amplitude level.}
\label{fig7}
\end{figure}


\section{Conclusion}

In this work we have introduced the 1D system which blends the $\mathcal{PT}$%
-symmetry, emulated SOC, and quintic (critical) nonlinearity. The system is
designed as a dual-core optical waveguide with skewed coupling between the
cores. The scheme makes it possible to produce families of stable solitons
in a very \textquotedblleft precarious" situation, as both the combination
of the gain and loss terms in the parallel waveguiding cores, which
represents the $\mathcal{PT}$-symmetry, and the critical self-focusing make
the solitons prone to the blowup instability. Nevertheless, the effective
SOC, which is represented by terms mixing the fields in the cores through
linear terms with the first spatial derivatives, added to the usual
(straight) linear coupling, turn out to be strong enough to stabilize parts
of the soliton families, which are found in both the main SIG (semi-infinite
gap) and the finite AG (annex gap), adjacent to the SIG. As concerns
unstable solitons, several scenarios of their evolution have been
identified. In addition to the generic blowup, weak instability may
transform the solitons into robust breathers. Interestingly, due to the
regularizing effect exerted by SOC on the system's instability beyond the
point of the breakup of the $\mathcal{PT}$ symmetry, stationary solitons are
found in this case too, although they are unstable. Simulations of
interactions between adjacent solitons have revealed both repulsion between
them and merger into a collapsing mode.

As an extension of the analysis, it may be relevant to consider effects of
spatial inhomogeneity, if it is present in the system. In particular, the $%
\mathcal{PT}$-symmetric gain-loss terms may be made inhomogeneous,
representing a spatially odd imaginary potential \cite{Bender}. It may be also
intersting to develop the analysis for systems with fractional diffraction
\cite{frac1,frac2}, and for systems including a trapping potential \cite%
{trapping,trapping2}, as well as for periodic waves \cite{periodic}. Another
possibility may be to apply machine-learning techniques to these systems
\cite{PINN}.

\section*{Declaration of Competing Interest}

The authors declare that they have no known competing financial interests or
personal relationships that could have appeared to influence the work
reported in this paper.

\section*{CRediT authorship contribution statement}

Gennadiy Burlak: Numerical simulations, Data analysis, Manuscript drafting.
Zhaopin Chen: Development of numerical methods, Numerical simulations, Data
analysis, Manuscript drafting. Boris A. Malomed: Conceptualization,
Analytical considerations, Data analysis, Manuscript drafting and editing.

\section*{Acknowledgments}

This work was supported, in part, by the Israel Science Foundation through
grant No. 1286/17 and, in part, by the CONACYT (M\'{e}xico) through grant
No. A1-S-9201. Z.C. acknowledges an fellowship provided by the Helen Diller
quantum center at the Technion (Haifa, Israel).

\end{document}